%% file: iml.tex
\documentclass[fleqn,usenatbib]{mnras}

\usepackage{newtxtext,newtxmath}
\usepackage[T1]{fontenc}
\usepackage{ae,aecompl}

\usepackage{graphicx}	% Including figure files
\usepackage{amsmath}	% Advanced maths commands
\usepackage{amssymb}	% Extra maths symbols

\usepackage{booktabs} %tables - toprule etc.

\usepackage[dvipsnames]{xcolor} % colors
\usepackage{subfig}

\newcommand{\angstrom}{\mbox{\normalfont\AA}}

\newcommand{\menu}{\href{https://toast-docs.readthedocs.io/en/latest/}{\texttt{toast-docs.readthedocs.io}}}

\newcommand{\gtoast}{\href{https://galaxyportal.space/}{\texttt{galaxyportal.space}}}

\newcommand{\srcd}{\href{https://github.com/ireis/portal/}{\texttt{github.com/ireis/portal}}}

\def\nad{\mbox{Na\hspace{.5pt}{I}\,D}}

\newcommand{\sklearn}{\texttt{scikit-learn}}
\newcommand\specfigwidth{1}

\title[UML in Astronomy]{Effectively using unsupervised machine learning in next generation astronomical surveys}

\author[I. Reis et al.]{
Itamar Reis$^{1}$\thanks{E-mail: itamarreis@mail.tau.ac.il},
Michael Rotman$^{2}$,
Dovi Poznanski$^{1}$,
J. Xavier Prochaska$^{3,4}$, \newauthor 
and Lior Wolf$^{2,5}$
\\
$^{1}$School of Physics and Astronomy, Tel-Aviv University, Tel-Aviv, 69978, Israel\\
$^{2}$School of Computer Science, Tel-Aviv University, Tel-Aviv, 69978, Israel\\
$^{3}$UCO/Lick Observatory, University of California, 1156 High Street, Santa Cruz, CA 95064, USA\\
$^{4}$Kavli Institute for the Physics and Mathematics of the Universe (WPI), University of Tokyo, Kashiwa 277-8583, Japan\\
$^{5}$Facebook AI Research
}

\date{Accepted XXX. Received YYY; in original form ZZZ}

\pubyear{---}

\begin{document}
\label{firstpage}
\pagerange{\pageref{firstpage}--\pageref{lastpage}}
\maketitle

\begin{abstract}
In recent years  many works have shown that unsupervised Machine Learning (ML)  can help detect unusual objects and uncover trends in large astronomical datasets, but a few challenges remain. We show here, for example, that different methods, or even small variations  of the same method, can produce significantly different outcomes.  While intuitively somewhat surprising, this can naturally occur when applying unsupervised ML to highly dimensional data, where there can be many reasonable yet different answers to the same question.  In such a case the  outcome of any single unsupervised ML method should be considered  a sample from  a conceivably wide range of possibilities. We therefore suggest an approach that eschews finding an optimal outcome,  instead facilitating the production and examination of many valid ones. This can be  achieved by incorporating unsupervised ML into data visualisation portals. We present here such a portal that we are developing, applied to the sample of SDSS spectra of galaxies. The main feature of the portal is interactive 2D maps of the data. Different maps are constructed by applying  dimensionality reduction to different subspaces of the data, so that each map contains different information that in turn gives a different perspective on the data.  The interactive maps are intuitive to use, and we demonstrate how peculiar objects and trends can be detected by means of a few button clicks. We believe that including tools in this spirit in  next generation astronomical surveys will be important for making unexpected discoveries, either by professional astronomers or by citizen scientists, and will generally enable the benefits of visual inspection even when dealing with very complex and extensive datasets. Our portal is available online at \gtoast.

\end{abstract}


%%%%%%%%%%%%%%%%%%%%%%%%%%%%%%%%%%%%%%%%%%%%%%%%%%

%%%%%%%%%%%%%%%%% BODY OF PAPER %%%%%%%%%%%%%%%%%%

\section{Introduction}

Every so often in the history of astronomy, visual inspection of data has led to an unexpected scientific discovery. As the volume of data that astronomical surveys gather increases, the presence of increasingly rare phenomena within these datasets is essentially unavoidable. It is worth noting that in astrophysics, what are observationally rare phenomena can actually be quite common and important but short lived (e.g., supernovae that only occur a handful of times per century in a given galaxy but play a crucial role in its chemical evolution). However, the likelihood  of detecting such rare phenomena is decreasing, since much of the data we now gather cannot realistically be visually inspected. 

The Sloan Digital Sky Survey \citep[SDSS; ][]{eisenstein11} has obtained spectra of $\sim 3$M galaxies and quasars. Since these were accumulated over nearly two decades, and intensively studied by the community, a non negligible fraction of the (high signal to noise ratio; SNR) data was in fact visually inspected by different people. This will surely not be the case for next generation surveys, that will generate $\sim10$ times more data in $\sim 1/10$ of the time. For example, the Dark Energy Spectroscopic Instrument \citep[DESI, ][]{levi13} which has begun commissioning, will observe $\sim 30$M galaxies, and SDSS-V \citep{kollmeier17} plan to obtain $\sim$6M stellar spectra and $\sim$25M spectra of the Milky Way interstellar medium (ISM). While the data rates increase a hundredfold, the number of experts available to inspect them remains more or less constant. 

Since we will not be able to examine every spectrum these surveys will generate,  we propose to enlist the help of unsupervised Machine Learning (ML) in order to prioritize. Unsupervised ML is the name of a broad family of tools that could be used to detect rare objects or trends. These tools include: (i) Clustering algorithms, which are used  to detect groups of objects sharing similar features, (ii) anomaly detection algorithms, which are used to detect unusual objects, and (iii) dimensionality reduction algorithms. The latter family of tools facilitates clustering, trend finding,  and anomaly-detection analyses, by representing a dataset in a low dimensional space that can be better visually (or otherwise) inspected. Since these algorithms are data rather than model driven, they have the potential to detect patterns  in the data that we did not know existed, and therefore would not have searched for directly.

A wide variety of approaches to perform each of these tasks has been developed by the ML community. \texttt{Python} implementations of many of the most common approaches are publicly available in \sklearn. These are relatively easy to use, with conveniently homogenized user interfaces. These tools  are  flexible and could be tuned to produce useful results from most datasets. Many examples of application of these methods to astronomical data are available in \texttt{AstroML} \citep{vanderplas12}.

In this work we make use of both anomaly detection and dimensionality reduction, in the following subsections we briefly review both,   focusing on recent usage in astronomy.

\subsection{Anomaly detection}
Anomaly detection algorithms typically rank objects based on a definition of abnormality, where many such definitions exist. Three common general approaches are: (i) objects that are the least similar to other objects in the data, (ii) objects that reside in low density regions of the data, and (iii) objects that are not well reconstructed by a model of the data. However, there are many ways to measure similarity, a variety of methods to define and measure the density, and obviously, endless approaches to modeling data, lending to the richness of existing anomaly detection algorithms. As could be seen in the examples below, all these approaches were successfully applied to astronomical datasets.

Applications of anomaly detection to spectroscopic data include \citet{bronson10} who detected anomalies in SDSS quasar spectra using the reconstruction error of a Principal Component Analysis (PCA) model of the data, i.e., the residual between the data and best fitting model. A more recent use of reconstruction-based anomaly detection is \citet{ichinohe19}, where the model of the data (in this case X-ray spectra), was built using a variational auto encoder. \citet{meusinger12} used self organizing maps \citep[SOM, ][]{kohonen82} for anomaly detection in SDSS quasar spectra. Their unusual quasars were defined to be objects residing in low density regions of the 2D embedding of the data created via SOM.  Distance based anomaly detection, was applied to SDSS galaxy and APOGEE stellar spectra by \citet{baron17a} and \citet{reis18} respectively, using an unsupervised Random Forest distance \citep{shi06}. 

For light curve data, distance based anomaly detection was used by \citet{protopapas06} and \citet{richards12} with different definitions of similarity. \citet{protopapas06} worked with raw light-curve data and used the cross correlation distance. \citet{richards12} worked with extracted features and used Random Forest (in this case it was supervised, trained on labeled data). In an example that does not directly fit into any of the general approaches described above, \citet{nun14} used a supervised Random Forest to predict the class of unlabeled objects, and anomalies were detected as objects having unusual voting distributions. \citet{nun16} detected anomalies using an ensemble of anomaly detection methods. 

An example of anomaly detection on imaging data is \citet{shamir14} who found peculiar SDSS galaxy pairs. As we discuss below, interpreting and understanding why an object was selected as anomalous by a given algorithm is often not trivial when dealing with spectroscopy or light curves. However, when inspecting the images of the objects detected in \citet{shamir14} their unusual features are manifest even to non experts in galaxy morphology.

\subsection{Dimensionality reduction}
Dimensionality reduction algorithms can be divided into types according to the quantity they try to preserve when representing the data in a lower dimensional space. Some algorithms, such as Multi Dimensional Scaling (\texttt{MDS}) try to preserve the similarity between all the objects. Other algorithms only try to only preserve the neighborhood, or nearest neighbors, but not the actual similarity values, e.g.,  \texttt{t-SNE} \citep{maaten08}. Auto-Encoders try to preserve information, in the sense that the data itself could be reproduced from the low dimensional representation, typically referred to as the latent space. In this work we use the Uniform Manifold Approximation and Projection (\texttt{UMAP}) algorithm \citep{umap}, which tries to preserve the topology of the manifolds on which the objects lie between the low dimensional representation and the original data. See \citet{gisbrecht15} for a review of dimensionality reduction algorithms. 

In recent years a number of works applied  dimensionality reduction techniques to astronomical datasets and showed that the resulting embeddings contain useful information \citep[][]{in-der-au12, meusinger12, jofre15, traven17, anders18, reis18}.  For example, \citet{reis18} created an  embedding of the APOGEE \citep{majewski16} infrared stellar spectra dataset using \texttt{t-SNE}. They showed that the location of a star on such a map contained information about its effective temperature,  surface gravity and  metallicity. In addition, groups of peculiar stars such as Be and carbon stars were clustered in specific locations on the map. 

While it is clear that such maps can contain non-trivial information, it is not obvious how we can extract this information and potentially learn something new about the data. We suggest, as also done by \citet{in-der-au12}, that one way to do this is  by using the maps in an interactive way. Selecting and inspecting objects directly from the map  enables studying the sample in detail. We will use such interactive maps of the data in the portal we present in Section \ref{sec:toast}.

\subsection{Outline}

In Section \ref{sec:uml} we  perform a comparison between various anomaly detection methods on a dataset of galaxy spectra from the SDSS.  From the method comparison we draw conclusions that are general to any unsupervised ML application, and we discuss a number of challenges to the effective incorporation of these methods into our workflow with future surveys. In Section \ref{sec:toast} we present the data portal that we are developing. We use unsupervised ML  to construct as many useful but different human-inspectable summaries of the data as possible, and gather them in an intuitive and easy to use interactive portal. We showcase this approach with SDSS galaxy spectra. We summarize in Section \ref{sec:sum}.

\section{The challenges with anomaly detection}
\label{sec:uml}

\subsection{Anomaly detection method comparison}
\label{sec:comparison}

The fact that many unrelated methods were successfully applied to astronomical data raises the question of how to choose which algorithm to use for a given project, or specifically, with next generation spectroscopic surveys. Is there, for a specific dataset, a single algorithm that is optimal?  To try to answer this question we use a sample of 150,000  SDSS galaxy spectra, and  apply four different anomaly detection algorithms; PCA reconstruction error, unsupervised Random Forest,  Fisher Vector based anomaly detection, and Isolation Forest.

For  PCA reconstruction error and Isolation Forest \citep{liu08} we use the  \sklearn{} implementation. For unsupervised Random Forest we use  the \sklearn{} implementation of Random Forest, and our own code (available at {\href{https://github.com/ireis/unsupervised-random-forest}{\texttt{github.com/ireis/unsupervised-random-forest}}) for calculating the anomaly score. We add to the comparison the results of the same algorithm  from \cite{baron17a}. For Fisher Vector based anomaly detection \citep{rotman19} we use our own implementation. The anomaly detection methods are described in more detail in Appendix \ref{app:methods}.
  
The galaxy spectra were obtained from the 14th data release of the SDSS \href{https://www.sdss.org/dr14/}{SDSS DR14}~\citep{abolfathi17}. We selected objects with \texttt{Class = GALAXY} from the \texttt{SpecObj} table and used only galaxies for which the rest frame spectrum contained flux values in the  wavelength range of $3700 \angstrom < \lambda  < 8000 \angstrom$. Out of these galaxies we selected the 150,000 with the highest SNR (according to the \texttt{SNMedian} field in the \texttt{SpecObj} table). 
The preprocessing stage consisted of removing flux values marked as bad by the SDSS pipeline (i.e., flux values with inverse variance of 0), normalizing the spectra by the median, shifting the spectra to the rest frame according to the SDSS pipeline redshift, and  interpolating  the spectra to a fixed wavelength grid. We note that, as shown for example in \citet{baron17a}, objects with incorrect redshifts can be found as outliers in this scheme. We use the normalized flux values as features for all the algorithms.

By definition, unsupervised tasks are challenging to optimize. What constitutes a successful application of anomaly detection? In science we typically aim to detect a large variety of anomalies, and the sole detection of objects of a single kind is considered non-satisfactory. In all cases significant tuning of the hyper-parameters or implementations (the difference between implementation decisions and hyper-parameters is  an  implementation decision) was required to obtain satisfactory results.

As an example of hyper-parameter tuning, with Isolation Forest we used rank values instead of the normalized flux values (that is, we strip the marginal distribution of each feature), to get satisfactory results. Without this modification we obtained only objects with extreme emission line strengths. Due to a small difference between the  \sklearn{} implementations of Random Forest and Isolation Forest, Random Forest is not sensitive to the marginal distributions of the features while Isolation Forest is. This is because in \sklearn{}, at each node of a Random Forest tree, the best split search grid is constructed from the feature values (of the objects in the node) themselves, while for an Isolation Forest tree the grid is a linearly spaced set of values, between the minimum and maximum feature values. This example illustrates  how seemingly insignificant implementation decisions can completely change the output of anomaly detection algorithms.  

Additional examples for implementation decisions and hyper-parameters  that we identified as having a major effect on the results include: (i) The properties of the synthetic data in Unsupervised Random Forest. This algorithm involves comparing the data to synthetic data that needs to be created by the user.  There is infinite freedom in constructing synthetic data, and naturally, this affects the results. We were able to obtain useful results with a number of different synthetic data types. See also \citet{shi06} who compared two relatively similar methods of creating synthetic data and obtained very different results; (ii) The number of objects used to train a single tree in Isolation Forest and Random Forest. This hyper-parameter has a major effect on the behavior of these algorithms, and yet it is not built into  \sklearn, and requires wrapping around their implementation; (iii) The number of components in the Gaussian mixture model for the Fisher Vector method; (iv) The details of the reconstruction error calculation for PCA reconstruction error. We were only able to obtain satisfactory results when calculating the reconstruction error separately on relatively small regions of the spectra, and inspecting objects with the largest errors in different regions.

\subsection{Consistency and significance of the discovered anomalies}

After tuning the different methods we were able to obtain satisfactory results with all four of them.  Considering all the methods uncovered a satisfactory list of anomalies, it is not trivial to pick the best one. Worst, we found that there is  little overlap between the anomalies detected  by the different methods and the mean overlap between the top $500$ objects from two different algorithms is $\sim 20$. The overlap between the anomalies detected with variations of the same method is also small, and in fact not different from the overlap between anomalies from completely different methods. Visually inspecting the detected anomalies showed that every method missed some interesting objects  found by others. This  suggests that a single anomaly detection algorithm does not produce in practice the most unusual objects in the data, rather the output should more appropriately  be considered as a small subset of the unusual objects. 

The overlap between the anomalies detected by  different methods can teach us about the statistical significance of the anomalies.  This is illustrated in Fig. \ref{fig:venns}. The top panels show Venn diagrams of two randomly drawn subgroups of size $g=500$, which were drawn from a parent population of size $n = 1,000$ (right) or size $n=10,000$ (left). The bottom panel show Venn diagram of  $g = 500$ anomalies detected by the two implementations of Random Forest, chosen here for example. The extent of the overlap between the two subgroups suggests that there are $n \sim 10,000$ anomalies in the data.

\begin{figure}

\includegraphics[width=\columnwidth]{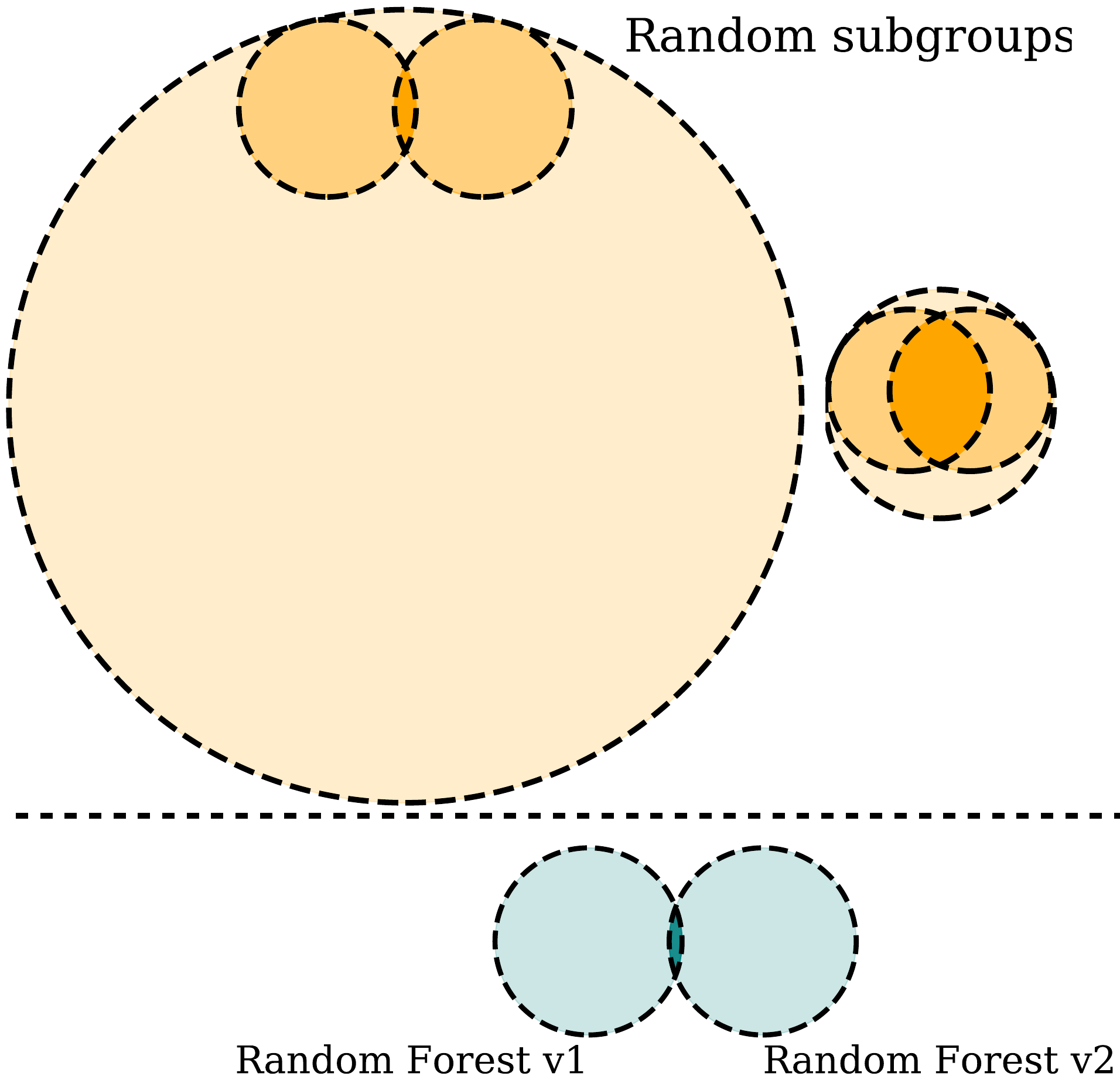}

\caption{An illustration of how the total number of anomalies in the data could be estimated from the overlap between the anomalies detected by different methods. The top  panels show  Venn diagrams of a group of a given size ($n=10,000$ on the left panel, and $n=1,000$ on the right panel) and 2 randomly drawn  subgroups of size $g = 500$. One can see the dependence of the overlap between the subgroups on the size of the original group. The bottom  panel show a Venn diagrams of  2  sets of 500  anomalies, detected by the two implementations of Random Forest. The overlap suggests we are in a situation similar to the one in the top left panel, where there is a large number of anomalies in the data, $\sim$ 10,000 in our case, of which we only detect small fractions.  A  quantitative estimation of the number of anomalies in the data is shown in Fig. \ref{fig:underlying_size}.} \label{fig:venns}
\end{figure}

This can be done quantitatively in the following way. Let us assume that there are $n$ anomalies in the data. Let $g$ be the number of anomalies detected by each method, and $k_{i,j}$ the overlap between the anomalies detected by methods $i$ and $j$. Further assuming that these sub-samples were uniformly drawn, we can  estimate  $n$ given $g$ and $k_{i,j}$, for each $i,j$.  To calculate the expected  $k$ given $n$ and $g$ consider having a group of objects of size $n$, and a subgroup of size $g$.  We are now randomly choosing, from the original group, another subgroup of size $g$. If $n \gg g$ the probability of a single randomly chosen object to be in the first subgroup is $g/n$. Choosing $g$ such objects,  the expected  overlap is $g^2/n$. In this case we can easily estimate the number of anomalies  with
\begin{equation}
    n \sim \frac{g^2}{k}.
\end{equation}
See Appendix \ref{app:fullcalc} for the calculation without assuming $n \gg g$. Given the formula, an intersect of $k= 20$ between $g=500$ anomalies detected by different methods suggests $n \sim 12,500$ anomalies in the data.

In Figure \ref{fig:underlying_size} we show the results of this calculation for each pair of anomaly detection methods we applied.  $k_{i,j}$  is shown on the x-axis, and the resulting $n$ is shown on the y-axis. The orange line represents  $n(k)$ for random groups and the shaded orange region is its standard deviation, calculated  numerically, by randomly drawing many groups and calculating the standard deviation of their intersects for different values of $n$. The largest overlap is between the PCA reconstruction and Fisher Vector methods (yellow-green line), for which $k_{i,j} = 29$ and $n = 8,525$. The lowest overlap is between one of the Random Forest implementations and PCA reconstruction (dark blue line), for this case $k_{i,j} = 8$ and $n = 31,716$. Note that the agreement between the two Random Forest implementation is  lower than that of completely different methods. While not shown in the figure, this is also true for different implementations of any of the other methods we applied. 

\begin{figure}
  \begin{center}
  \includegraphics[width=\columnwidth]{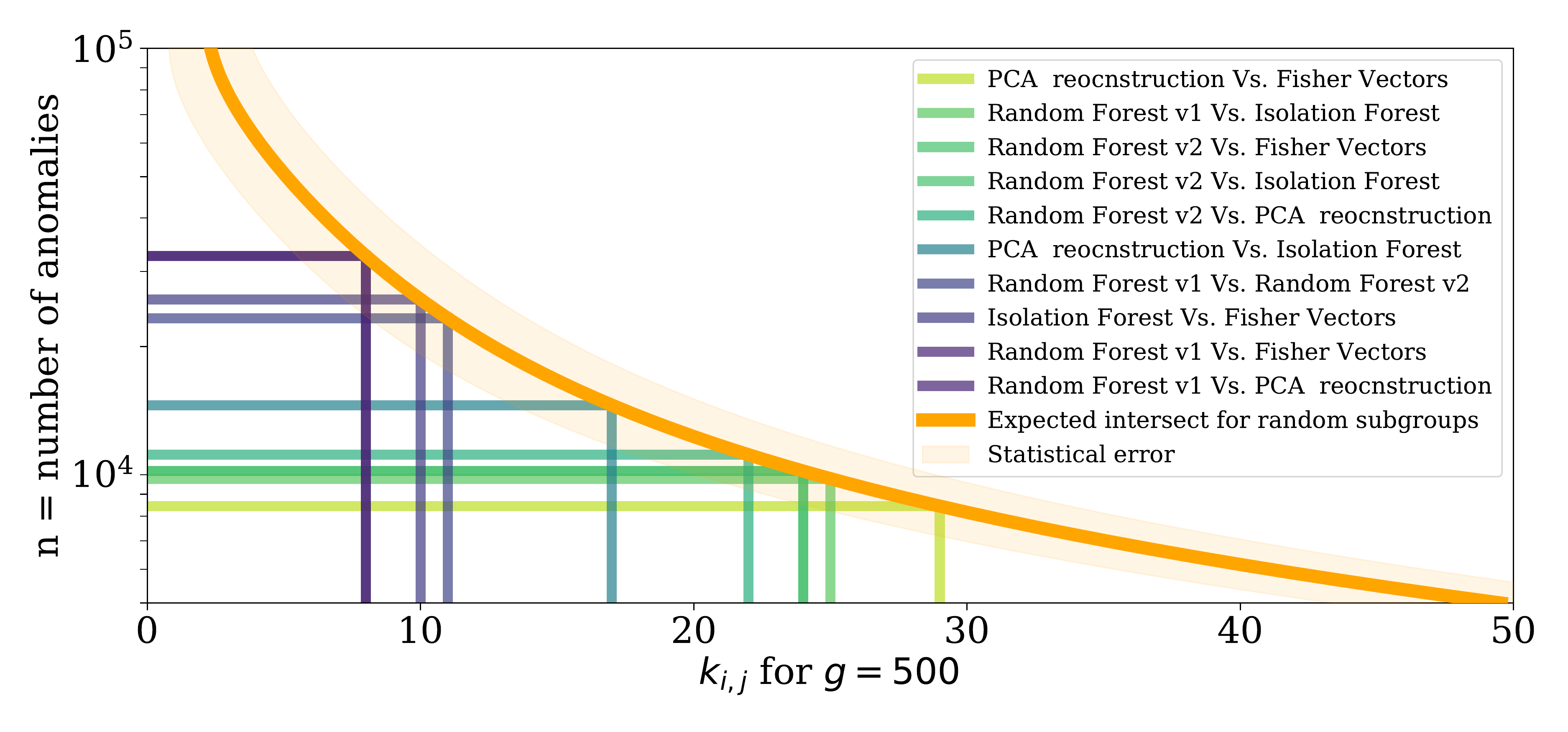}
  \caption{The overlap ($k_{i,j}$, x-axis) between the top $g=500$ detected anomalies of different anomaly detection methods, and the resulting estimated total number of anomalies in the data ($n$, y-axis). The orange line is the expected size of a parent population given an overlap between two randomly drawn subgroups of size 500. If one assumes simplistically that there is a well defined group of anomalies from which different detection method pick  random sub-samples, one can measure the size of the underlying group of anomalies. Using it only as a ballpark estimate, it seems that the underlying group is of order $10^4$, which is a few percent of the data. Random Forest v1 and v2 are two implementations of the same algorithm.  The statistical uncertainty in the expected intersect is shown in light orange, while the Poisson uncertainty is omitted. }\label{fig:underlying_size}
  \end{center}
\end{figure}

Taking this result at face value means that our dataset contains $\mathcal{O}(10^4)$ anomalies that would be picked up by either of these algorithms. As a consequence, by choosing only one algorithm, even if optimized, and visually inspecting a few hundred candidates, a common and manageable number, we will only assess  $\mathcal{O}(5\%)$ of the true anomalies.  This fraction is expected to further decrease as the number of objects worth noting will increase with the size of the data, but the number of objects inspected by a single person is not. Next generation spectroscopic surveys will contain one or two orders of magnitude more objects than what we have considered here, thus requiring a different approach. 

For the sake of the discussion above we have made the assumption that each anomaly detection method detects a random group of anomalies. If this assumption were true, then the distribution of the anomalies detected by any single method would have been representative of the overall distribution of anomalies in the data. In such a case each single method would have the potential to detect all types of anomalies.  This would be surprising, and indeed does not hold in practice, as can be seen for example in Figure \ref{fig:anomalies_snr}, where we show the SNR distribution of the top 100 anomalies from three methods. The fact that the three distributions are so different from each other makes it clear that the different methods are biased towards different types of objects. Note that if two methods are biased towards the same types of objects their intersect will be larger than that of randomly drawn groups and decrease our estimate of the number of anomalies in the data, and vice versa. The biased nature of the detected anomalies should thus increase the scatter in Figure \ref{fig:underlying_size} but not necessarily change the mean value. To get an unbiased sample, that has a better chance of detecting all types of anomalies, it is necessary to use an ensemble of  different methods. This however does not obviate the need to examine many more objects than typically done. 

Also, we used 500 objects per algorithm, but our results above are largely insensitive to that choice, in that as long as we use a few hundred objects per algorithm, we consistently get $n = \mathcal{O} (10^4)$. With more than a few hundreds, $n$ starts to increase, likely due to false positives. Using less than $\sim200$ objects per algorithm creates zero overlaps between the subsamples (and brings the difficulties of small number statistics).

\begin{figure}
  \begin{center}
  \includegraphics[width=\columnwidth]{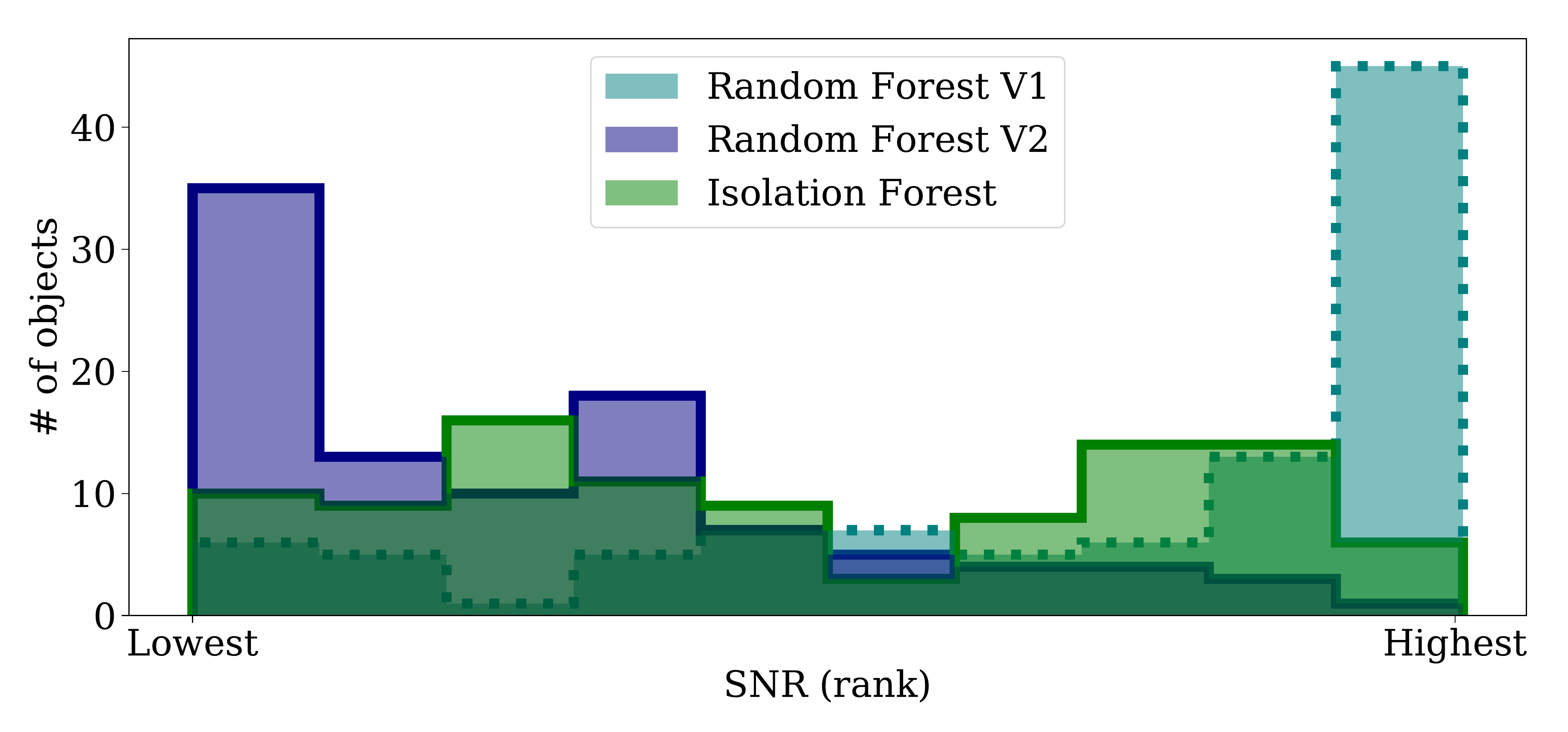}
  \caption{The Signal to Noise Ratio (SNR) distribution of the top 100 anomalies from 2 implementations of unsupervised Random Forest, and an implementation of Isolation Forest, on the SDSS galaxy spectra dataset. While one implementation of unsupervised Random Forest favors high SNR objects, the other favors low SNR object. This Isolation Forest implementation does not seem to show any SNR preference. This demonstrates how different detection methods or specific implementations are biased differently and produce inherently different outcomes. The x-axis is the SNR rank and not the SNR value, otherwise the SNR distribution dominates the graph.}\label{fig:anomalies_snr}
  \end{center}
\end{figure}

\subsection{Interpretability of the anomalies}

Interpreting  anomalies can be an even greater challenge than detecting them. All anomaly detection methods discussed above only produce a list of unusual objects and do not provide any indication as to what is unusual about a given object. This critical task is very time consuming, and prone to errors as there is  no guarantee that our interpretation will be correct, i.e., that we indeed found the reason that an object was tagged as anomalous. 

Interpretation becomes even more challenging when a single object is too complex to be easily inspected in a glance. This would be the case for spatially resolved spectroscopy, spectroscopic time series, or high resolution spectra, to give examples from the world of spectroscopy. As the complexity of the data increases, the question of what is unusual about a given object becomes more challenging but also more interesting. Furthermore, with the complexity,  more objects will show some unusual features, or as often coined in the ML literature: when the complexity of the data is high enough every object is an anomaly. All the anomaly detection methods we have discussed are inherently designed for low dimensional data in which the anomalies are obvious and only need to be quickly and automatically detected. Progress will therefore come from a more streamlined interpretation, rather than more clever detection algorithms. 

\subsection{Unsupervised ML for high dimensional data}
\label{sec:highd}

The challenges  we discussed are both stemming from  the complexity (or high dimensionality) of datasets and not their size. These challenges  are not special to astronomy, and in fact anomaly detection in high dimensional data is   an active ML field of research   \citep[][listing a few examples]{aggarwal01, zhang04a, muller08, kriegel09a, muller10, keller12}.  See also \citet{zimek12} for a ML oriented review of this topic.

Before discussing possible approaches for handling high dimensional data we would like to emphasize  that while we  focus on anomaly detection,  other unsupervised ML applications, such as dimensionality reduction and clustering, suffer from  the same issues. For example, in high dimensional data there could be many different but informative ways to divide  objects into clusters, and different clustering algorithms can produce different but valid outcomes. This is illustrated in Figure \ref{fig:clstr} for the case of data with only three independent features. We see that these data can be divided into clusters in a number of different ways, without any single way being the best one.

\begin{figure}
  \begin{center}
  \includegraphics[width=\columnwidth]{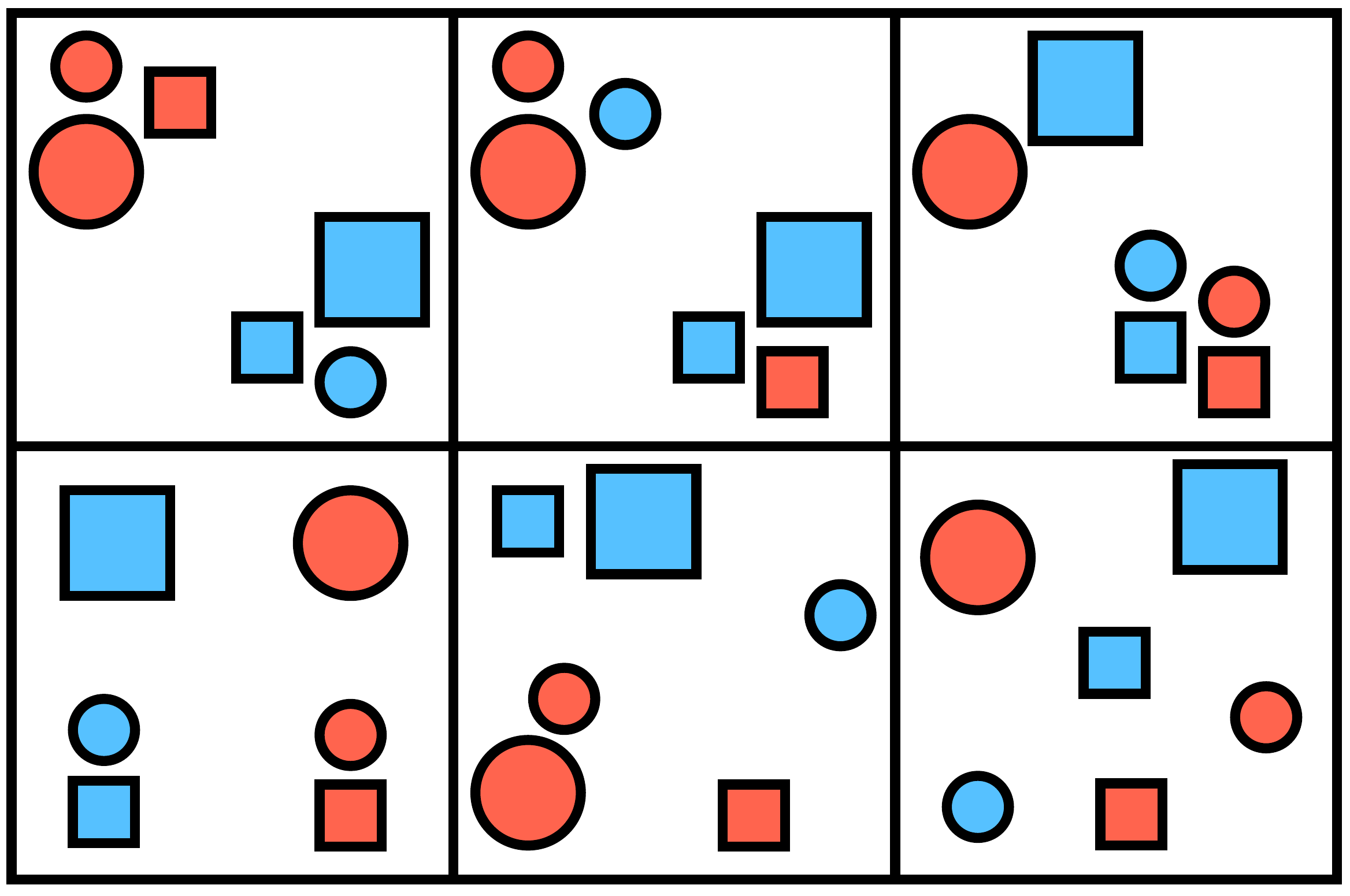}
  \caption{Six different  ways  of clustering the same data, none of which can be qualified as optimal. As the complexity of data increases, and each object is represented with more features, there could be many ways to divide the data into clusters (or similarly to define anomalies). While different from each other, all are conceivable. This is an illustration of the fact that in unsupervised ML there could be many possible answers to the questions we are asking. We should not look for the best answer, but  instead find and inspect as many valid ones as possible.} \label{fig:clstr}
  \end{center}
\end{figure}

The reason that anomaly detection, clustering, and dimensionality reduction all suffer from the same issues, could be reduced to the definition of a pair-wise distance between objects in the data, which is the basis of many unsupervised ML methods. In complex data the relationship between two objects cannot be described by a single distance. If two objects are similar in one feature and different in another,  there is no good way to include  the information in a single number. As exemplified in Figure \ref{fig:clstr}, is a red square more similar (i.e., closer) to a red circle or a blue square? Different definition of distance will give more weight to some features and less to others, and thus produce different outcomes for any unsupervised ML application based on this distance. Each different outcome has the potential of providing useful information. 

A common approach for handling high dimensional data is working with subspaces, each subspace containing all the objects, but only a subset of the features. The advantage of this approach is that with small enough subspaces the issues arising due to the high dimensionality of the data disappear. In a small subspace the definition of distance becomes unique so there will be a single way to cluster the data, and  define  anomalies. The results we obtain are also  easier to interpret, as we know why an object is an anomaly, or what is the common feature of cluster members, according to the subspace in which the anomaly or cluster were detected. 

By working with subspaces one can resolve the two issues discussed above. One can scan many solutions, each from a given algorithm in a given subspace with its own merit, and with an easier path to interpretation, since an outlier in a limited subspace would be easier to understand, as we indeed show below. However, the unavoidable cost of having many algorithms, run within many interesting subspaces, is that there are many results to vet, more than can be realistically done by a single human.

\section{Framework for the exploration of unsupervised ML results}
\label{sec:toast}

One  way to address the difficulty of having many different but useful outcomes, is collecting them in an easy to use tool so that they could be inspected by the community. In order to allow large numbers of people from the community to explore current and future datasets through the unsupervised ML lens, we are building an interactive graphical portal  to large and complex datasets. 

Data exploration portals are common in astronomy, a few examples are: \texttt{SIMBAD} \citep{wenger00},  \texttt{Galaxy Zoo} \citep{lintott08}, The Open Supernova Catalog \citep{guillochon17}, \texttt{ESASky} \citep{baines17}, \texttt{Marvin} \citep{cherinka18},   and \texttt{SkyPortal} \citep{walt19}. An upcoming portal for anomaly  detection in astronomical data is \texttt{Astronomaly} \citep{lochner}. These portals are generally designed to inspect single objects in a convenient way, and  thus cannot be used to explore large numbers of sources. 

The key novel feature of our portal is  machine-learned  two-dimensional embeddings (or maps) of the data, that automatically group sources which are similar to each other, and from which objects can be interactively selected and inspected. The interactive maps should be used with the simple notion that similar objects  are located close to each other.     This makes detecting potentially interesting phenomena quite intuitive. Objects that are isolated on a map are interpreted as objects that are not similar to any other object in the dataset. Any structure in the ordering of objects suggests a continuous change in the properties of objects along the structure. A compact group of objects implies the objects share some common properties. While the maps are built using ML their usage is intuitive and does not require any prior knowledge in this field. Furthermore, by working in subspaces of the data, we can produce a number of such maps, each containing different information.  As discussed above, when working with  such subspaces, the interpretation of  any unsupervised ML method, and specifically the 2D maps,  is also made easier. 

 To illustrate the possible use cases of such a portal we apply it to the SDSS galaxy spectra dataset (same dataset as described in Section \ref{sec:comparison}). In the next subsection we discuss the details of our portal as currently implemented for this dataset. A screenshot of the portal is presented in Figure \ref{fig:screen}. The portal itself is available online at \gtoast{}, a user manual could be found at \menu{}. The source code is publicly available at \srcd{}.

\begin{figure}
  \begin{center}
  \includegraphics[width=\columnwidth]{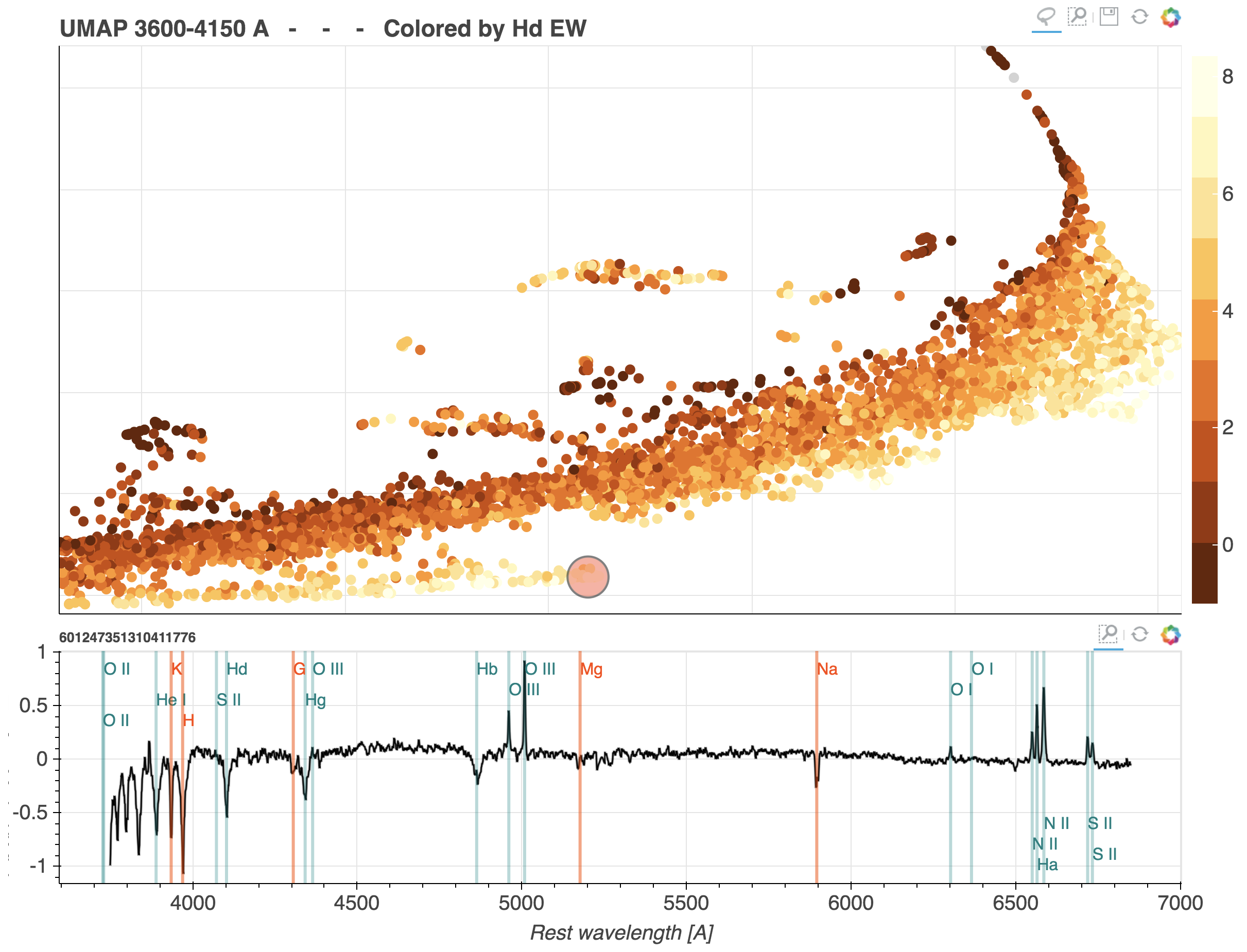}
  \caption{A screenshot from our data portal (\gtoast{}), where one can study the SDSS galaxy spectra through an unsupervised ML lens. In the top panel, each point represents a galaxy, and galaxies with similar spectral properties  are located close to each other in this abstract plane. In this specific embedding the similarities  are based on the wavelength region  $3600 - 4150 [\angstrom]$. The map is further colored by the equivalent width of the H$_{\delta}$ absorption line. This map and coloring are just examples of the many that we made available. The user of the portal can easily toggle between various embeddings and coloring schemes. In the bottom panel we see the spectrum of the galaxy that has been selected interactively in the top panel, where it is marked with a red circle. 
  } \label{fig:screen}
  \end{center}
\end{figure}

\subsection{Implementation details}
\label{sec:2de}

Our implementation for the SDSS galaxies is intended to showcase the general approach described above, and  the details can change in other implementations or in future versions of this portal. The main challenge in constructing the portal was creating  interactive linked graphs containing large amounts of data. Since showing hundreds of thousands of points on a graph is both prohibitive and pointless, we implemented an adaptive graph that shows a random subset (with a fixed  upper limit size) of the objects in the current frame, where more objects in a specific region could be seen by zooming in. Our code is based on the \texttt{Bokeh}  library\footnote{https://bokeh.pydata.org/en/latest/}.

To create the maps of the data we first divide the spectra to a number of wavelength regions which are manually defined according  to the locations of common emission and absorption lines.  These wavelength regions  are examples of subspaces of the data. In each region the spectrum is normalized by the median flux value in the region. Next, Euclidean distances are calculated between the objects (i.e., a sum of the squared differences of normalized fluxes). Finally, we apply the \texttt{UMAP} algorithm to the distances to obtain the maps. 

To illustrate the advantages of using maps created in subspaces of the data, let us consider a map constructed from the region of the spectrum containing the \nad\ doublet. An isolated group of galaxies on this map will most likely contain galaxies with unique \nad\ line profiles. On the other hand, on a map constructed from the entire spectrum, (i) there is no guarantee the same group of galaxies with unique \nad\ profile will appear as an isolated group, since on such a map the galaxies could be grouped according to some other more prominent feature, and (ii) given an isolated group on the map it is hard to determine what is the unique feature shared by the objects in the group. Figure \ref{fig:outs_umap} shows an example of the first point, where objects clustered on a map created using the \nad\ region are no longer clustered on a map constructed from a different wavelength region.

In addition to the machine learned maps, we uploaded to the portal common galaxy diagnostics such as the  BPT diagrams \citep{bpt} from which objects can be interactively selected as well. All the embeddings in the portal are linked which allow the user to select objects on one map and display them on another. With this, it is easy to check, for example, where galaxies with unusual \nad\  profiles lie on the BPT diagram. This example is shown in the bottom left panel of Figure \ref{fig:outs_umap}. 

With the map displaying 2D information, it is easy to add a third one via color. The maps can be colored by various properties of the galaxies such as line ratios,   the star formation rate, and  the velocity dispersion. Most properties are taken from the SDSS value added catalogs, the rest were calculated by us. Coloring the maps is useful when learning the general location of different types of galaxies on the map. For example see Figure \ref{fig:screen} where the \texttt{UMAP} is colored by the H$_{\delta}$ equivalent width (EW). In this example the coloring can guide the user to the locations of post-starburst galaxies. The bottom panel of Figure \ref{fig:screen} shows one such post-starburst galaxy, located at the end of a one dimensional structure with increasing H$_{\delta}$ EW.

We included in the portal the results of all the anomaly detection algorithms that we used for the method comparison in section \ref{sec:comparison}. The location of the anomalies on the different maps is useful for investigating the reason a given object was detected as an anomaly.  It is also possible to order the anomalies according to either  their location on the map, or  any of the available galaxy properties, instead of viewing them ordered by their abnormality score, which is effectively quite random. This allows for a more efficient visual inspection process, as similar objects can be inspected and classified together.

Since anomaly detection is only one of the goals of our portal, we also include other features, that can be used when searching for trends. For example, the user can bin and stack spectra of objects on the fly, by selecting them on the maps, and  binning them according to their coordinates  or  any of the available galaxy properties. A few example use cases for these features are presented below, more extensive instructions on how to use all the features are available in the online documentation at \menu{}.

\subsection{Use case I: Anomaly detection}

\label{sec:nad_anomalies}

As a demonstration of the capabilities of the portal for anomaly detection, we focus here on a single map, constructed from the $\lambda = 5680 - 6120 [\angstrom]$ region. As we show below, this restricted view alone uncovers multiple interesting phenomena. 
 We detect galaxies as anomalies by manually inspecting the map. Namely we  select and inspect all objects that are either isolated or located at extreme ends of the 2D distribution of objects. The map showing the detected groups is presented in Figure \ref{fig:outs_umap} and a number of example objects are shown in Figure \ref{fig:outs}. A full description of the findings is given below.

\begin{figure*}

\includegraphics[width=\textwidth]{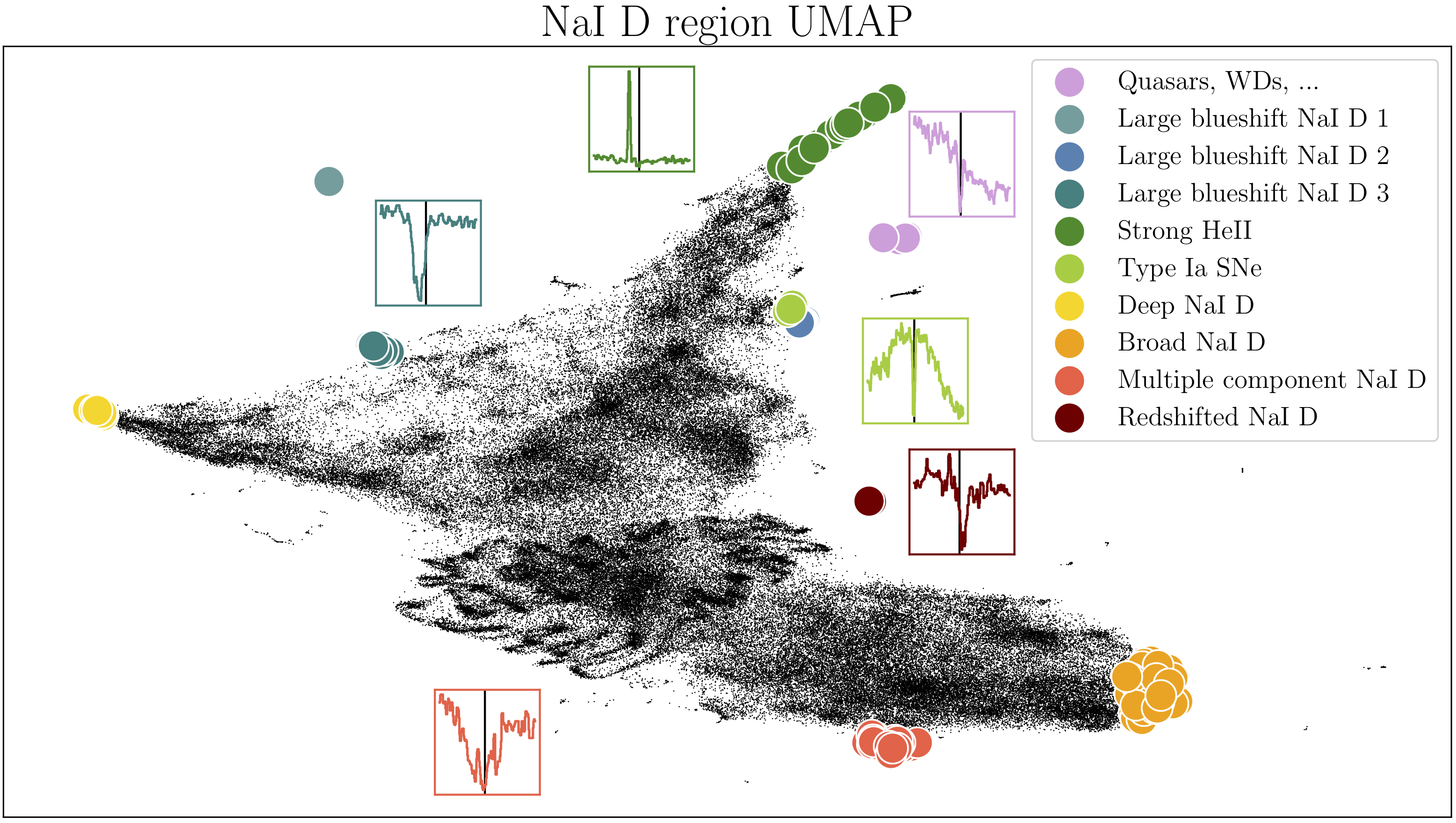}  \\[6pt]
\includegraphics[width=\columnwidth]{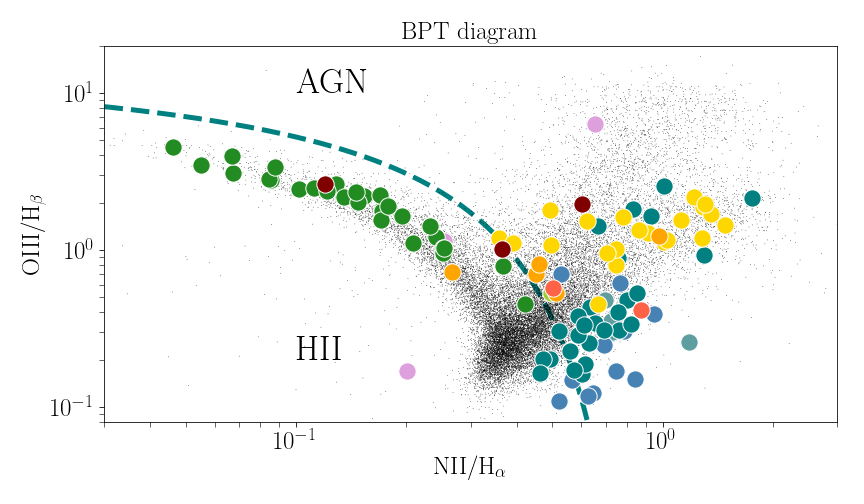}   \includegraphics[width=\columnwidth]{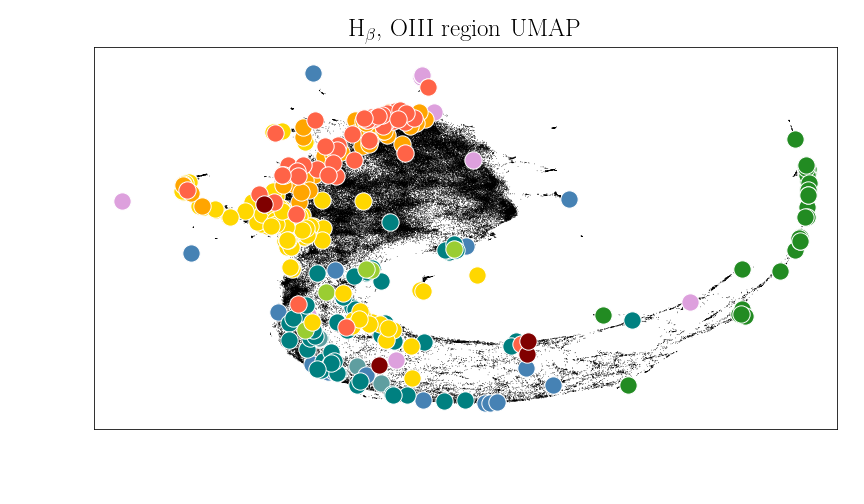}

\caption{\textbf{Top panel:} A \texttt{UMAP} of SDSS galaxies constructed from the $\lambda = 5680 - 6120 [\angstrom]$ region of the spectrum. Groups of unusual objects detected from the map are shown in color. Inserts containing the \nad\ part of the spectrum for an example object are shown for select groups. \textbf{Bottoms panels:} The locations of the same galaxies with unusual \nad\ properties as in the top panel, on different embeddings. The bottom left panel shows the BPT diagram which separates AGNs from star-forming galaxies. The bottom right panel shows a \texttt{UMAP}  constructed from the $\lambda = 4700 - 5100 [\angstrom]$ region of the spectrum, containing the H$_{\beta}$ and OIII lines.  In the bottom left panel we can see that some of the unusual \nad\ groups lie in specific regions of the BPT diagram. The interesting finding is the galaxies with deep \nad\ absorption, who seem to be preferentially located in the AGN region of the BPT, while many galaxies with strong blueshifted \nad seem to be outliers on the BPT diagram.  In the bottom right panel we see that the unusual \nad\ objects could not have been detected using this $\lambda = 4700 - 5100 [\angstrom]$ map, illustrating that different maps contain different information. The locations of the unusual \nad\ objects here are also not random, suggesting correlations between various \nad\ and H$_{\beta}$ and OIII properties.
 Interactive versions of all these maps, from which the unusual \nad\ objects and others  were selected and inspected, is available at \gtoast.} \label{fig:outs_umap}
\end{figure*}

In this wavelength region, the strongest features are the \nad\ absorption doublet (to which both  cold stars and the ISM contribute), and the HeII emission line. We find the following groups of unusual objects (for the more interesting groups we list a few examples in a table, more examples can be obtained from the portal itself):

\begin{itemize}
\item As could be expected, galaxies with extremely strong \nad\ absorption are located at an extreme end of the 2D distribution and are easily detected. They are shown in yellow in Figure \ref{fig:outs_umap}. 

\item Similarly expected, galaxies with extreme HeII emission are found at another  edge. They are shown in dark green. 

\item Galaxies with strongly blueshifted \nad\ are grouped together at a few locations. We mark them in blue.  The largest concentration contains a few hundred galaxies, where the objects with the largest blueshifts are also the farthest from the bulk of the galaxies.   Another small group of galaxies with blueshifted \nad\ contains a few galaxies in which the absorption  is also stronger. The last group of these is characterized by a bluer continuum. Table \ref{tab:blueshift} lists a number of examples. 
 Blueshifted \nad\ is a signature of outflowing gas. A sizable fraction of the objects we find are star-forming face-on spirals such as the ones discussed in \citep{heckman02, bae18}. Another dominant population is composed of post-starburst galaxies. These are found using the portal by inspecting the locations of the strongly blueshifted \nad\ galaxies on the \texttt{UMAP} constructed from a region of the spectrum containing the H$_{\delta}$ absorption, where post-starburst galaxies cluster, or by coloring the \nad\ map with the H$_{\delta}$ EW. 

\item Galaxies with multiple component \nad\ are located in a single cluster on the map. They are shown in light red. Most galaxies in this group seem to be well described by two velocity components, but some show evidence of three. The images of the objects in this group all show that these objects are blended, suggesting the multiple components are coming from different galaxies at similar redshifts. Some of these objects appear in catalogs of galaxy clusters. Table \ref{tab:multi} lists a number of  examples.

\item We find a number galaxies with redshifted \nad\ absorption. Four such objects  are located in a small cluster on the map. These are shown in dark red  in Figure \ref{fig:outs_umap}. Three additional objects are found in the region of the map containing galaxies with strong \nad{} absorption (shown in yellow on the map). All these objects are listed in Table \ref{tab:redshifted}. Inspecting the individual objects, it seems that different reasons cause the apparent redshifted \nad\ line. In \href{http://skyserver.sdss.org/dr14/en/tools/explore/summary.aspx?sid=4718658695541530624&apid=}{SDSS J211635.95-004613.1}, the emission lines are redshifted by the same velocity as the \nad\ line, while the $H$ and $K$ lines are centered on the systematic redshift. On the other hand, in \href{http://skyserver.sdss.org/dr14/en/tools/explore/summary.aspx?sid=3135796793174943744&apid=}{SDSS J141518.01+230841.0}, the $H$ and $K$ line are redshfited with the \nad\ line, while the emission lines are blueshifted from the systematic redshift. In the second case a natural interpretation is an offset AGN \citep[e.g,][]{comerford14}.

\item Another cluster on the map is composed of galaxies with an unusual slopes of their SED in this wavelength region. They are shown in purple. Many of these objects  were mistakenly classified as galaxies by the SDSS pipeline but are in fact white dwarfs and quasars. The objects that are indeed galaxies, such as \href{http://skyserver.sdss.org/dr14/en/tools/explore/summary.aspx?sid=863651577993390080&apid=}{SDSS J092034.87+511224.1} and \href{http://skyserver.sdss.org/dr14/en/tools/explore/summary.aspx?sid=863649928725948416&apid=}{SDSS J091959.69+513346.7} show  broad emission features that we think are due to some error in the data acquisition or reduction.

\item Galaxies hosting type Ia supernova features in their spectra are also found in a small cluster. We mark them in light green. 8 of these galaxies, already found  by \citet{graur13}, are listed in \ref{tab:supernova}. Additional candidates could be found by inspecting the rest of the objects in the cluster.  The clustering of these galaxies  is  due to the fact that the tell-tale feature of these supernovae, the deep Si absorption, falls partly within this wavelength region. Maps constructed with different wavelength regions, in which there are no prominent supernova spectral features, do not contain such a cluster.

\item Additional clusters (not shown in color) and objects that are isolated on the map show a variety of unusual behaviors. We find a small  cluster of objects with chance alignment with brown dwarfs, objects with the wrong redshift determined by the SDSS pipeline, objects with bad sky lines subtraction, and some with unexplained absorption lines (likely due to a foreground or background source).  Some of these galaxies are listed in Table \ref{tab:isolated}.

\end{itemize}

The bottom left panel of Figure \ref{fig:outs_umap} shows the locations of the detected anomalies on the BPT diagram. Note that some of the anomalous galaxies do not have detected emission lines and are thus not shown on the diagram. 3 groups of unusual \nad\ region galaxies also cluster on the BPT diagram: (i) Objects with strong \nad\ absorption are located in the AGN region of the diagram. (ii) Many of the objects with strongly blueshifted \nad\ are outliers in the BPT diagram, and have unusually low OIII/H$_{\beta}$ line ratios given their (high) NII/H$_{\alpha}$ ratios. (iii) Objects with strong HeII emission are located in the star forming branch of the diagram.

To illustrate that different maps contain different information we show the locations of the same unusual groups of objects discussed above on a \texttt{UMAP} constructed from a different wavelength region in the bottom right panel of Figure \ref{fig:outs_umap}. This map shows the $\lambda = 4700 - 5100 [\angstrom]$ region, containing the H$_{\beta}$ and OIII lines. The unusual \nad\  groups are no longer clustered or isolated and would not be detected on this map.

\iffalse
\begin{figure}

\includegraphics[width=\columnwidth]{bpt_test.pdf}  \\[6pt] \includegraphics[width=\columnwidth]{hb_oiii_umap_test.pdf}

\caption{The locations of galaxies with unusual \nad\ properties on different embeddings. The top panel shows the BPT diagram. The bottom panel shows a \texttt{UMAP}  constructed from the $\lambda = 4700 - 5100 [\angstrom]$ region of the spectrum, containing the H$_{\beta}$ and OIII lines. On both panels the same groups of unusual objects detected from the \nad\ map (Figure \ref{fig:outs_umap}) are shown in color. On the top panels we can see that some of the unusual \nad\ groups lie on specific regions of the BPT diagram. The more interesting cases are those of galaxies with deep \nad\ absorption, located in the AGN region of the BPT, and galaxies with strong blueshift \nad\ that lie on the a relatively under-populated region of the BPT.  In the second panel we see that the unusual \nad\ objects could not have been detected using this $\lambda = 4700 - 5100 [\angstrom]$ map, illustrating that different maps contain different information. The locations of the unusual \nad\ objects are also not random, suggesting correlations between various \nad\ and H$_{\beta}$ and OIII properties. } \label{fig:outs_umap_other_wave}
\end{figure}
\fi

\subsection{Use case II: Trend detection}

One method to detect trends is by inspecting how objects change along structures on the maps.  In this example we use the $\lambda = 6400 - 6700 [\angstrom]$ region map. This region contains the H$_{\alpha}$ and NII lines. Objects that lie along geometrical structures on the map are expected to show continuous changes in these features. The user of our portal can select a specific region on the map and a specific direction, and stack the galaxies in bins along the chosen direction. This procedure takes a few button clicks on the portal. Figure \ref{fig:stacks} shows an example. What is found in this case is that the galaxies are ordered according to a combination of the emission line width and the H$_{\alpha}$ to NII amplitude ratio. In general the emission features are more  AGN-like towards the stack colored in yellow. Trends can be discovered by  looking for correlated changes in other regions of the spectrum. With well known samples or object types, one will mostly find trivial trends (e.g.,  emission line amplitudes that are correlated). In this example, however, we find that the equivalent width of the \nad\ absorption correlates with the width of the H$_{\alpha}$ -NII complex. We see that the more AGN-like stacks  have stronger \nad\ absorption. Note that the bottom left panel of Figure \ref{fig:outs_umap}  shows a similar trend as objects with deep \nad\ absorption lie in the AGN region of the BPT diagram.

\begin{figure}
  \begin{center}
  \includegraphics[width=\columnwidth]{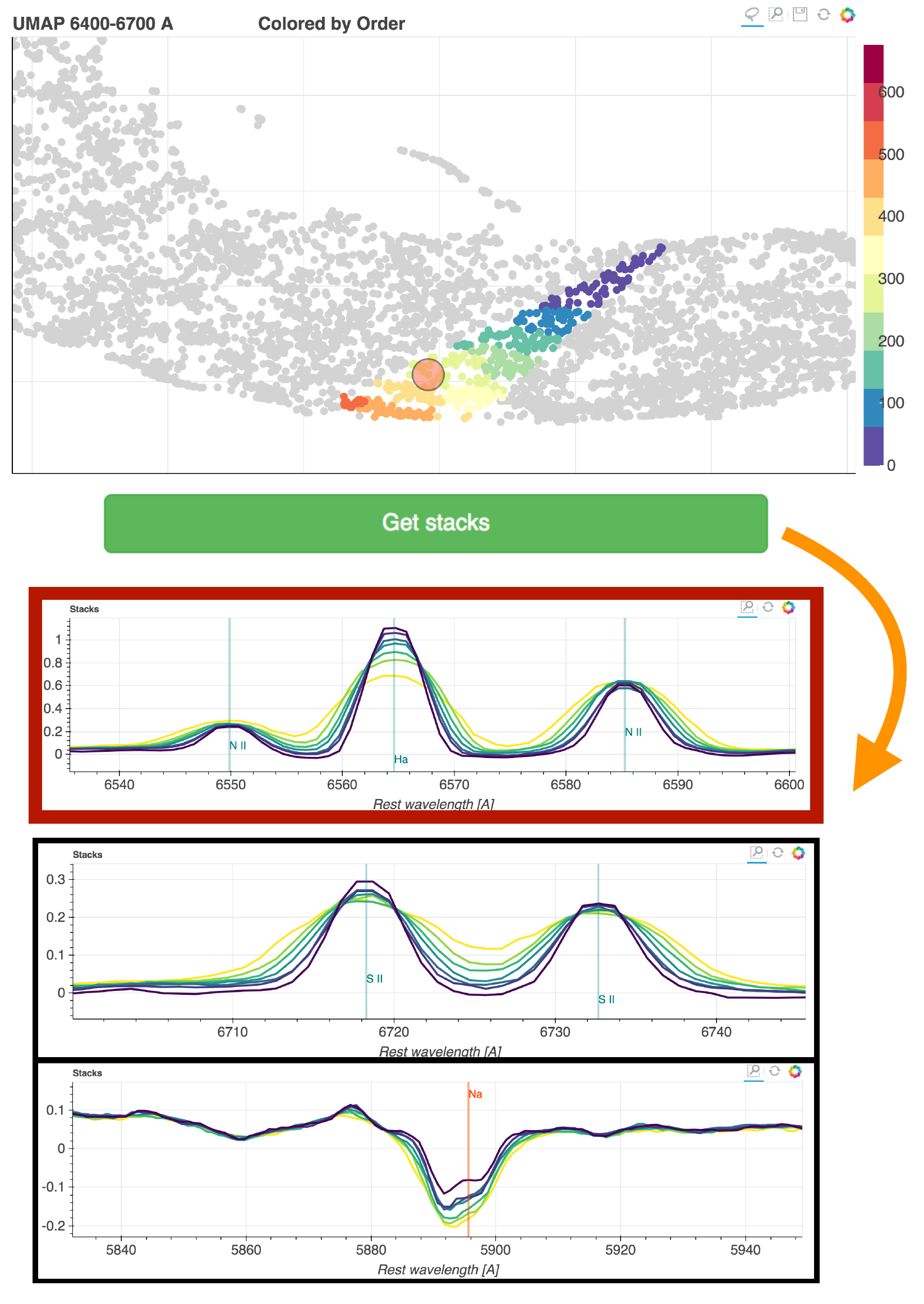}
  \caption{An example of trend exploration. The top panel shows a part of the $\lambda = 6400 - 6700 [\angstrom]$  map, from which some of the galaxies are selected  (the ones that are colored). The galaxies are ordered according to their location on the map, with color following the order. The bottom panels show different wavelength regions of the stacked spectra of the galaxies binned according to this order. In the example shown here the galaxies seem ordered by a combination of the line width and line ratio. The yellow lines have more AGN-like emission features; broader emission lines and higher  NII to H$_{\alpha}$ line ratio. In this example  we also  see a non-trivial correlation with the \nad\ absorption line that seems stronger in the AGN-like galaxies.}\label{fig:stacks}
  \end{center}
\end{figure}

\section{Summary}
\label{sec:sum}

We apply various  anomaly detection methods to the same dataset of SDSS galaxy spectra,  and show that while they all succeed, they disagree on most of the top few hundred outliers. This naively surprising result is a natural manifestation of the fact that for high dimensional data there could be many  different yet reasonable answers to the questions we ask of any unsupervised ML algorithm. As a consequence, any single method will only output  a small subset of the answers we wish for.

In order to increase our chances of making discoveries, a practical approach is to accumulate many different results of unsupervised ML and make them available for inspection by the community in an  easy and intuitive way.  For this, we are developing an exploration tool for astronomical data, that brings together interactivity and unsupervised ML. The main feature of the portal is 2D embeddings, or maps, of the data, where objects that are similar in a given subspace are located near each other.  

We demonstrate and develop our portal using the dataset of SDSS galaxy spectra. For this dataset we built several maps, each constructed using a different wavelength region. We also include several additional embedding (e.g, the BPT diagrams), various coloring schemes using metadata, and the results of a number of anomaly detection methods. 

We show how such tools and approach can allow for a more effective discovery process when interested in anomaly detection or when searching for trends and correlations. Additional uses include the search for objects of interest via similarity (`find me more objects like this one'; \citealt{reis18a}), or for studying an object in context of its peers (`how similar is this objects to others from other perspectives?'). We believe such methods will soon be indispensable.

\section*{Acknowledgements}
This research was supported by Grants No. 2014413 and 2018017 from the United States-Israel Binational Science Foundation (BSF), and Grant No. 541/17 from the Israeli Science Foundation (ISF). 
JXP acknowledges support to the Applied Artificial Intelligence Initiative by UC Santa Cruz.
This research made use of:
{\fontfamily{cmtt}\selectfont Bokeh},  {\fontfamily{cmtt}\selectfont  scikit-learn} \citep[][]{pedregosa11}, {\fontfamily{cmtt}\selectfont  SciPy} \citep[including {\fontfamily{cmtt}\selectfont  pandas} and {\fontfamily{cmtt}\selectfont NumPy, }][]{scipy01},  {\fontfamily{cmtt}\selectfont IPython} \citep[][]{perez07}, {\fontfamily{cmtt}\selectfont matplotlib} \citep[][]{hunter07}, {\fontfamily{cmtt}\selectfont Astropy} \citep[][]{astropy-collaboration13}, {\fontfamily{cmtt}\selectfont Numba} \citep{lam15},  the SIMBAD database \citep[][]{wenger00}, and the NASA Astrophysics Data System Bibliographic Services.

This work made extensive use of SDSS data. Funding for the Sloan Digital Sky Survey IV has been provided by the Alfred P. Sloan Foundation, the U.S. Department of Energy Office of Science, and the Participating Institutions. SDSS-IV acknowledges
support and resources from the Center for High-Performance Computing at
the University of Utah. The SDSS web site is www.sdss.org.

SDSS-IV is managed by the Astrophysical Research Consortium for the 
Participating Institutions of the SDSS Collaboration including the 
Brazilian Participation Group, the Carnegie Institution for Science, 
Carnegie Mellon University, the Chilean Participation Group, the French Participation Group, Harvard-Smithsonian Center for Astrophysics, 
Instituto de Astrof\'isica de Canarias, The Johns Hopkins University, 
Kavli Institute for the Physics and Mathematics of the Universe (IPMU) / 
University of Tokyo, Lawrence Berkeley National Laboratory, 
Leibniz Institut f\"ur Astrophysik Potsdam (AIP),  
Max-Planck-Institut f\"ur Astronomie (MPIA Heidelberg), 
Max-Planck-Institut f\"ur Astrophysik (MPA Garching), 
Max-Planck-Institut f\"ur Extraterrestrische Physik (MPE), 
National Astronomical Observatories of China, New Mexico State University, 
New York University, University of Notre Dame, 
Observat\'ario Nacional / MCTI, The Ohio State University, 
Pennsylvania State University, Shanghai Astronomical Observatory, 
United Kingdom Participation Group,
Universidad Nacional Aut\'onoma de M\'exico, University of Arizona, 
University of Colorado Boulder, University of Oxford, University of Portsmouth, 
University of Utah, University of Virginia, University of Washington, University of Wisconsin, 
Vanderbilt University, and Yale University.

%%%%%%%%%%%%%%%%%%%%%%%%%%%%%%%%%%%%%%%%%%%%%%%%%%



%\bibliographystyle{mnras}

\bibliographystyle{mnras}
\bibliography{iml}

%%%%%%%%%%%%%%%%%%%%%%%%%%%%%%%%%%%%%%%%%%%%%%%%%%

%%%%%%%%%%%%%%%%% APPENDICES %%%%%%%%%%%%%%%%%%%%%

\appendix

\section{Number of anomalies given the overlap}
\label{app:fullcalc}
To calculate the probability of obtaining an overlap of $k$ without assuming $n \gg g$, consider a specific order of randomly choosing the objects, where the first three objects are in  the subgroup, and the next two are not. The probability for this scenario is 
\begin{equation}
    \frac{g}{n} \times \frac{g-1}{n-1} \times \frac{g-2}{n-2} \times \frac{n-g}{n-3} \times \frac{n-g-1}{n-4}.
\end{equation}
We see that for any order in which the objects are chosen, the probability for choosing $k$ objects form the first subgroup and $g-k$ from the rest of the group is 
\begin{equation}
    \frac{g(g-1)\dots(g-k)\times(n-g)(n-g-1)\dots(n-g - (g-k))}{n (n-1)  (n-2) \dots  (n-g)} .
\end{equation}

Multiplying by the number of possible orders to choose the objects, the probability to obtain an overlap of $k$ can be written as 
\begin{equation}
    \frac{g!}{(g-k)!} \frac{(n-g)!}{((n-g) - (g-k) )!} \frac{(n-g)!}{n!}
    {g\choose k}.
\end{equation}

\section{Anomaly detection methods}
\label{app:methods}
In this section we briefly describe the four anomaly detection methods that we used in the method comparison, applied to the galaxy spectra, and uploaded to our portal. 

\begin{enumerate}

\item PCA reconstruction error. This is an example for model-based approach for anomaly detection. PCA  is used to model the data, and the anomaly score is defined to be the $\chi^2$  between the model and the data. Additional examples of ways to model the data that could be used instead of PCA include independent component analysis (ICA), non-zero matrix factorization (NMF), Auto-Encoder,  and physically motivated models.

\item Unsupervised Random Forest. This is an example for distance based approach for anomaly detection in which the definition of distance definition is based on an ensemble of classification trees. The similarity between two objects is defined to be the number of trees in which the objects end up on the same terminal node when propagated through the tree. The distance is the inverse of the similarity. An unsupervised Random Forest, which is a Random Forest \citep{breiman84} trained to distinguish between real and synthetic data, is an example for an ensemble of classification trees. Other  tree ensembles that could be used with the same distance definition include supervised Random Forest, Extremely Randomized Trees (ERTs), and boosted trees \citep[e.g AdaBoost,][]{freund97}. Many other distance definitions exist, the simplest one being Euclidean distance.

\item Isolation Forest. This algorithm does not fall into any of the general three approaches given in Section \ref{sec:uml}. In Isolation Forest the inverse of the anomaly score is defined to be the number of nodes an object goes through before being isolated (i.e, found to be the only object on a node), summed over all the trees in  the ensemble. Isolation Forest is commonly used with ERTs, but in principle could be used with other types of classification tree ensembles.

\item Fisher Vector based anomaly detection. This algorithm is related to the density based approach for anomaly detection. Instead of using the density itself, the anomaly score is defined to be the contribution of an object to the Fisher Information the data holds about the parameters of the density distribution model. We  used a Gaussian Mixture Model (GMM) to model the density of the data, as in this case the gradients of the density with respect to its parameters, which are needed for the calculation of the anomaly score, have an analytical formula. 

\end{enumerate}

\section{\nad\ anomalies specific examples}
\label{app:anomalies_tables}
In this section we provide examples from the various types of galaxies detected as anomalies from the \nad\ \texttt{UMAP}, as described in Section \ref{sec:nad_anomalies}. Figure \ref{fig:outs} shows example spectra of a number of different types of such anomalies.  For each object the left panel shows the entire SDSS spectrum, and the right panel shows the zoom in on the \nad{} region which was used to detect the objects. A number of example objects for the different types of detected anomalies listed in the following tables: Table \ref{tab:blueshift}: galaxies with large blueshift \nad{}. Table \ref{tab:multi}: galaxies with multiple component \nad{}. Table \ref{tab:redshifted}: galaxies with redshifted \nad{}. Table \ref{tab:supernova}: galaxies hosting a type Ia supernova. Table \ref{tab:isolated}: Various additional types of anomalies. 

\begin{figure*}

\subfloat[\href{http://skyserver.sdss.org/dr14/en/tools/explore/summary.aspx?sid=728629081251276800&apid=}{SDSS J234028.01-090945.0} -  \nad\ absorption is redshifted relative to the emission lines.]{\includegraphics[width=\specfigwidth\columnwidth]{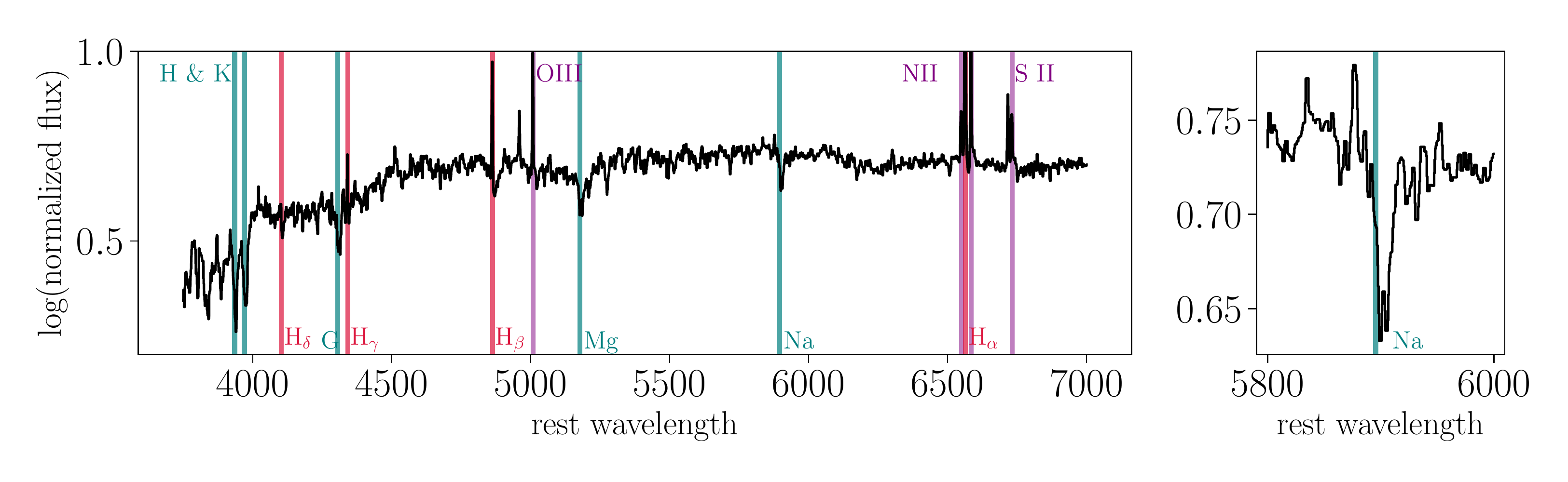}  }
\subfloat[\href{http://skyserver.sdss.org/dr14/en/tools/explore/summary.aspx?sid=683536182147049472&apid=}{SDSS J142812.98+611115.6} - extreme blueshift \nad.]{\includegraphics[width=\specfigwidth\columnwidth]{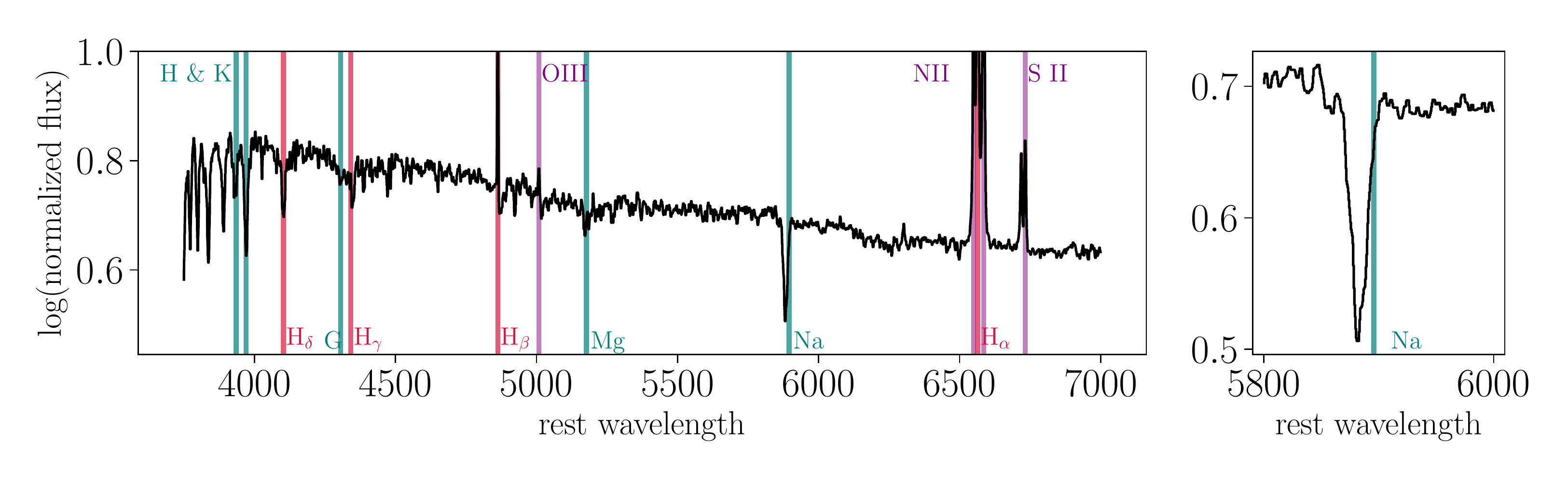} } \\[6pt]
\subfloat[\href{http://skyserver.sdss.org/dr14/en/tools/explore/summary.aspx?sid=308647239662725120&apid=}{SDSS J104230.55+003441.9} - \nad\ in emission along with blueshifted absorption. ]{\includegraphics[width=\specfigwidth\columnwidth]{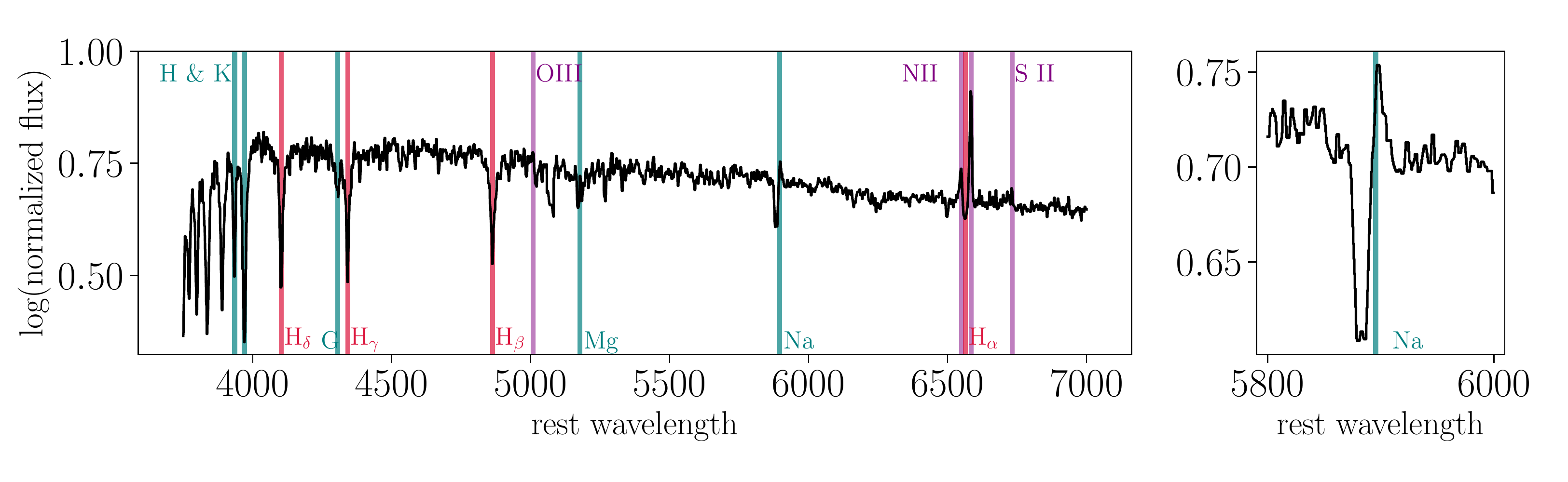} } 
\subfloat[\href{http://skyserver.sdss.org/dr14/en/tools/explore/summary.aspx?sid=2674019488826943488&apid=}{SDSS J103219.59+194052.6} - multiple component \nad\ absorption.]{\includegraphics[width=\specfigwidth\columnwidth]{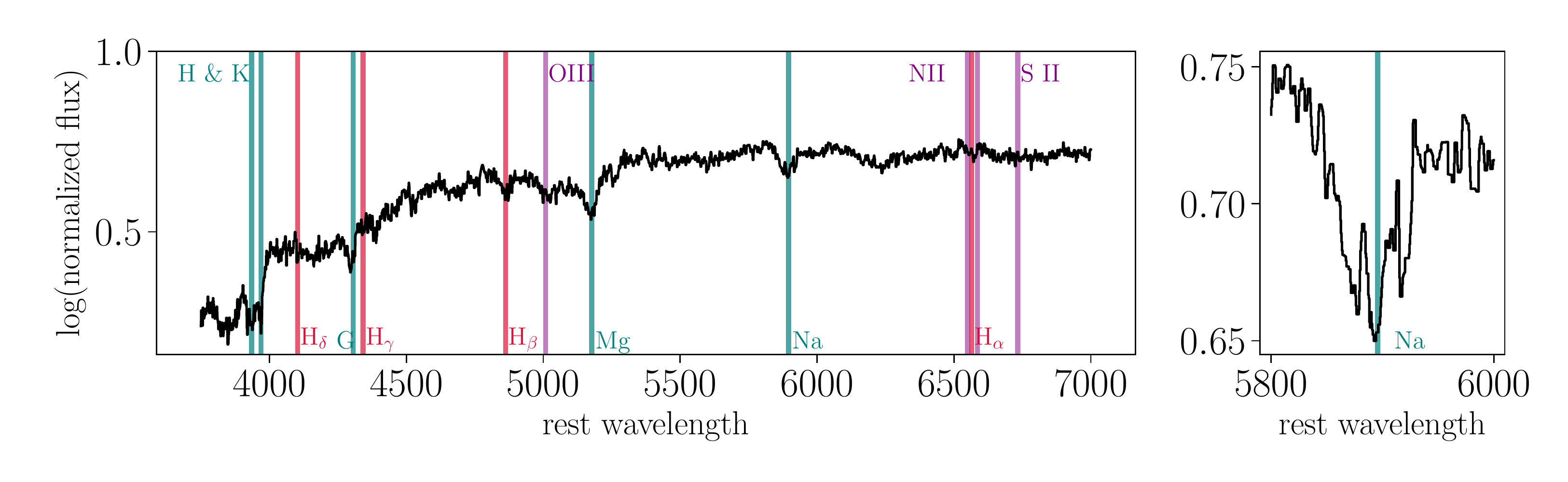}} \\[6pt]

\subfloat[\href{http://skyserver.sdss.org/dr14/en/tools/explore/summary.aspx?sid=6534893589381488640&apid=}{SDSS J091337.33+295958.4} - type Ia supernova. ]{\includegraphics[width=\specfigwidth\columnwidth]{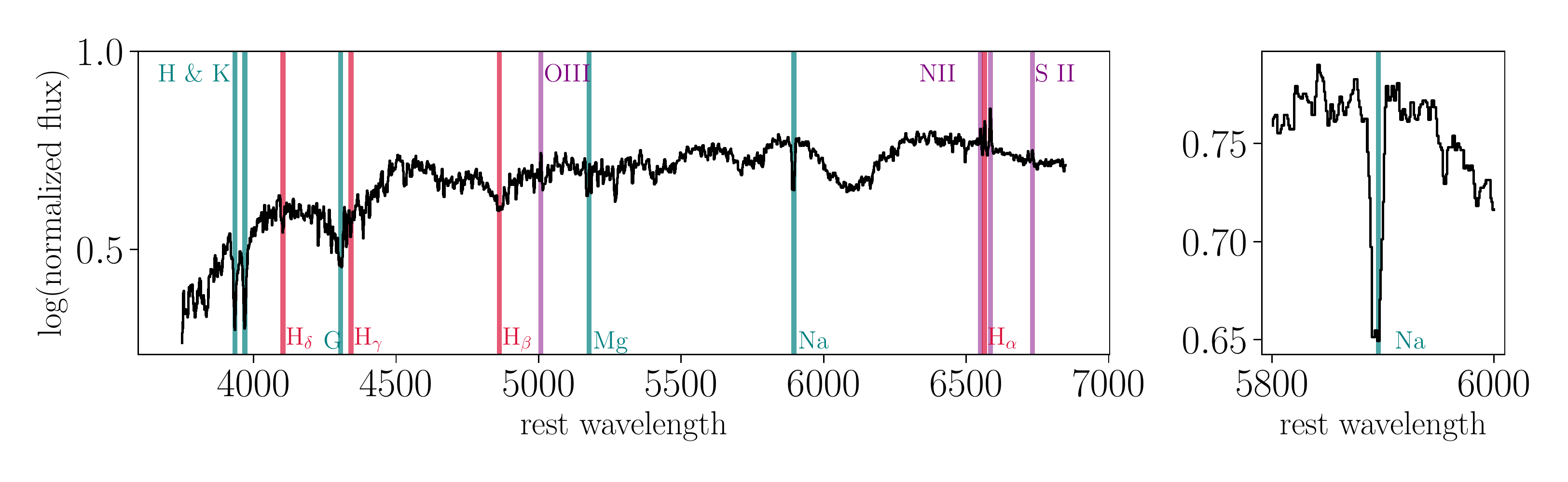} } 
\subfloat[\href{http://skyserver.sdss.org/dr14/en/tools/explore/summary.aspx?sid=863651577993390080&apid=}{SDSS J092034.87+511224.1} - unidentified broad feature.]{\includegraphics[width=\specfigwidth\columnwidth]{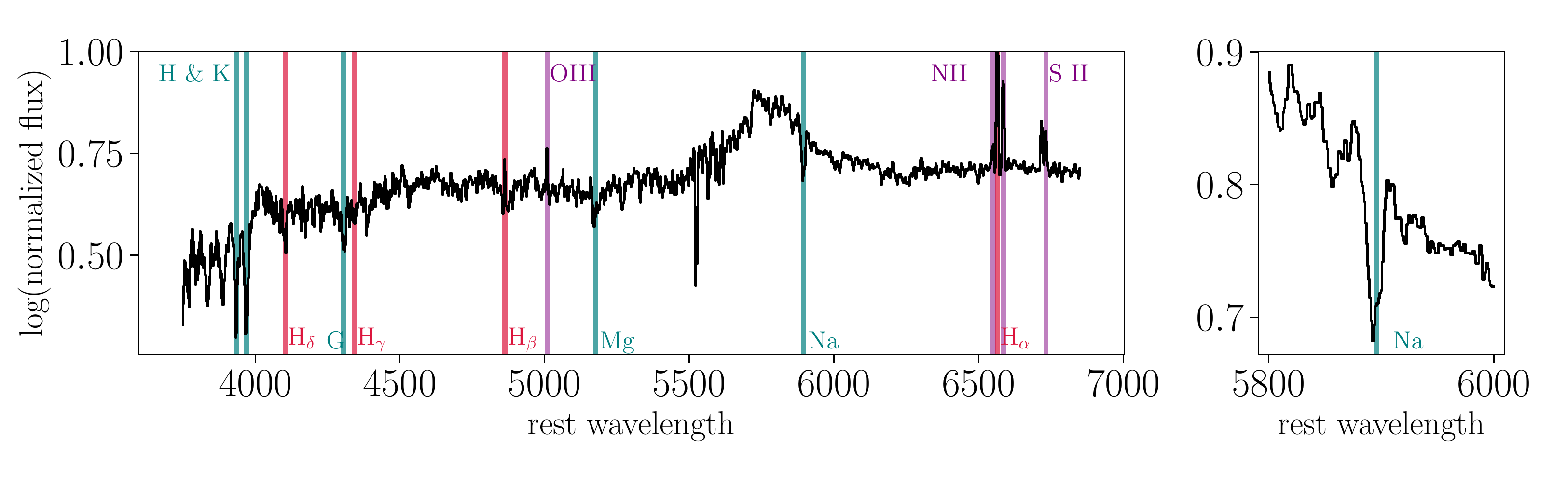}} \\[6pt]

\caption{Examples of galaxies with unusual \nad\ line profiles detected with our portal.   The blue vertical lines mark the locations of common absorption lines, the purple vertical lines mark common emission lines, and the red vertical lines mark the Balmer series that can be seen in both absorption and emission.} \label{fig:outs}
\end{figure*}

\input{blueshift.txt}

\input{multi.txt}

\input{redshifted.txt}

\input{supernova.txt}

\input{isolated.txt}

%%%%%%%%%%%%%%%%%%%%%%%%%%%%%%%%%%%%%%%%%%%%%%%%%%

% Don't change these lines
\bsp	% typesetting comment
\label{lastpage}
\end{document}

%% file: blueshift.txt
\newcounter{magicrownumbers}
\newcommand\rownumber{\stepcounter{magicrownumbers}\arabic{magicrownumbers}}

\begin{table}
\begin{center}

\tiny
\begin{tabular}{l|ccc}

\toprule
index & SDSS name  & comments     \\
\midrule
\rownumber &  \href{http://skyserver.sdss.org/dr14/en/tools/explore/summary.aspx?sid=683536182147049472&apid=}{SDSS J142812.98+611115.6} &  SF  \\

%\rownumber &  \href{http://skyserver.sdss.org/dr14/en/tools/explore/summary.aspx?sid=1794693572675528704&apid=}{SDSS J094630.90+345500.6} &  wolf-rayet galaxy, SF  \\

\rownumber &  \href{http://skyserver.sdss.org/dr14/en/tools/explore/summary.aspx?sid=1794693572675528704&apid=}{SDSS J094630.90+345500.6} &  wolf-rayet galaxy, SF, FOS  \\

\rownumber &  \href{http://skyserver.sdss.org/dr14/en/tools/explore/summary.aspx?sid=308647239662725120&apid=}{SDSS J104230.55+003441.9} &  E+A  \\

\rownumber &  \href{http://skyserver.sdss.org/dr14/en/tools/explore/summary.aspx?sid=609140194892867584&apid=}{SDSS J073856.16+320317.4} &  SF,  E+A  \\

\rownumber &  \href{http://skyserver.sdss.org/dr14/en/tools/explore/summary.aspx?sid=588906983042607104&apid=}{SDSS J125427.34+022059.3} &  SF   \\

\rownumber &  \href{http://skyserver.sdss.org/dr14/en/tools/explore/summary.aspx?sid=2391510971186178048&apid=}{SDSS J140621.04+252846.9} &  SB, ionized outflows   \\

\rownumber &  \href{http://skyserver.sdss.org/dr14/en/tools/explore/summary.aspx?sid=2673058790458288128&apid=}{SDSS J125427.34+022059.3} &  E+A   \\

\rownumber &  \href{http://skyserver.sdss.org/dr14/en/tools/explore/summary.aspx?sid=1831919745926981632&apid=}{SDSS J122715.39+062757.2} &  SF, E+A   \\

\rownumber &  \href{http://skyserver.sdss.org/dr14/en/tools/explore/summary.aspx?sid=3525236668162482176&apid=}{SDSS J235047.12+143617.5} &  SF, ionized outflows   \\

\rownumber & \href{http://skyserver.sdss.org/dr14/en/tools/explore/summary.aspx?sid=1882639117586556928&apid=}{SDSS J141943.23+491411.9} &           SF, FOS \\

 \rownumber &
 \href{http://skyserver.sdss.org/dr14/en/tools/explore/summary.aspx?sid=2346477991690790912&apid=}{SDSS J083950.75+230836.1} &           SF, FOS \\
 \rownumber &
 \href{http://skyserver.sdss.org/dr14/en/tools/explore/summary.aspx?sid=4793025814357581824&apid=}{SDSS J025600.55+013829.5} &           SF, FOS \\
 \rownumber &
 \href{http://skyserver.sdss.org/dr14/en/tools/explore/summary.aspx?sid=464003560229595136&apid==}{SDSS J031034.09+002938.7} &           SF, FOS \\

 \bottomrule
\end{tabular}
\caption{Examples for galaxies with extreme \nad{} blueshifts. Additional features are included in the comments column: SF = star forming emission lines, SB = star burst, FOS = face on spiral.}
\label{tab:blueshift}

\large
\end{center}
\end{table}

%% file: multi.txt
%\newcounter{magicrownumbers}
%\newcommand\rownumber{\stepcounter{magicrownumbers}\arabic{magicrownumbers}}

\setcounter{magicrownumbers}{0}

\begin{table}
\begin{center}

\tiny
\begin{tabular}{l|ccc}

\toprule
index & SDSS name  & number of components     \\
\midrule
\rownumber &  \href{http://skyserver.sdss.org/dr14/en/tools/explore/summary.aspx?sid=4239226546928197632&apid=}{SDSS J084344.28+385340.6} &  2 \\

\rownumber &  
\href{http://skyserver.sdss.org/dr14/en/tools/explore/summary.aspx?sid=4591501823613837312&apid=}{SDSS J211138.95+044126.8} &  2 \\

\rownumber & 
\href{http://skyserver.sdss.org/dr14/en/tools/explore/summary.aspx?sid=2674019488826943488&apid=}{SDSS J103219.59+194052.6} &  3 \\
 
\rownumber & 
\href{http://skyserver.sdss.org/dr14/en/tools/explore/summary.aspx?sid=1618052082776958976&apid=}{SDSS J110534.89+410524.2} &  2 \\

\rownumber & 
\href{http://skyserver.sdss.org/dr14/en/tools/explore/summary.aspx?sid=2272156235715340288&apid=}{SDSS J125856.36+385053.4} &  2 \\

 \bottomrule
\end{tabular}
\caption{Galaxies with multi component \nad{} profile.}
\label{tab:multi}

\large
\end{center}
\end{table}

%% file: redshifted.txt
%\newcounter{magicrownumbers}
%\newcommand\rownumber{\stepcounter{magicrownumbers}\arabic{magicrownumbers}}

\setcounter{magicrownumbers}{0}

\begin{table}
\begin{center}

\tiny
\begin{tabular}{l|ccc}

\toprule
index & SDSS name  & absorption & emission     \\
\midrule
\rownumber &  \href{http://skyserver.sdss.org/dr14/en/tools/explore/summary.aspx?sid=2445543179315865600&apid=}{SDSS J155157.88+203056.9} &  redshifted & blueshifted  \\

\rownumber &  \href{http://skyserver.sdss.org/dr14/en/tools/explore/summary.aspx?sid=728629081251276800&apid=}{SDSS J234028.01-090945.0} &  redshifted & blueshifted  \\

\rownumber &  \href{http://skyserver.sdss.org/dr14/en/tools/explore/summary.aspx?sid=1091119921124894720&apid=}{SDSS J120525.71+510611.1} &  redshifted & multi-component  \\

\rownumber &  \href{http://skyserver.sdss.org/dr14/en/tools/explore/summary.aspx?sid=1484036179434170368&apid=}{SDSS J125553.16+581948.6} &  redshifted & multi-component  \\

 \rownumber &  \href{http://skyserver.sdss.org/dr14/en/tools/explore/summary.aspx?sid=443768422623373312&apid=}{SDSS J005555.93+003940.2} &  multi-component & redshifted  \\
 
  \rownumber &  \href{http://skyserver.sdss.org/dr14/en/tools/explore/summary.aspx?sid=3135796793174943744&apid=}{SDSS J141518.01+230841.0} &  redshifted & blueshifted  \\
  
\rownumber &  \href{http://skyserver.sdss.org/dr14/en/tools/explore/summary.aspx?sid=4718658695541530624&apid=}{SDSS J211635.95-004613.1} &  multi-component & redshifted  \\

 \bottomrule
\end{tabular}
\caption{Galaxies showing redshifted \nad{} absorption. The absorption and emission columns refers to the location of the features relative to the SDSS systematic redshift. The absorption column refers to the other absorption lines in the spectrum, e.g, the $H$ and $K$ lines.}
\label{tab:redshifted}

\large
\end{center}
\end{table}

%% file: supernova.txt
%\newcounter{magicrownumbers}
%\newcommand\rownumber{\stepcounter{magicrownumbers}\arabic{magicrownumbers}}

\setcounter{magicrownumbers}{0}

\begin{table}
\begin{center}

\tiny
\begin{tabular}{l|ccc}

\toprule
index & SDSS name       \\
\midrule

\rownumber &  \href{http://skyserver.sdss.org/dr14/en/tools/explore/summary.aspx?sid=5342400104622956544&apid=}{SDSS J080821.09+005035.3}  \\

\rownumber & 
\href{http://skyserver.sdss.org/dr14/en/tools/explore/summary.aspx?sid=3091895762661959680&apid=}{SDSS J142608.24+152501.9}  \\

\rownumber & 
\href{http://skyserver.sdss.org/dr14/en/tools/explore/summary.aspx?sid=540438585724135424&apid=}{SDSS J095153.06+010605.8}  \\

\rownumber & 
\href{http://skyserver.sdss.org/dr14/en/tools/explore/summary.aspx?sid=2999526070589351936&apid=}{SDSS J132301.39+243023.6}  \\

\rownumber & 
\href{http://skyserver.sdss.org/dr14/en/tools/explore/summary.aspx?sid=2035656513569187840&apid=}{SDSS J140309.73+060754.3}  \\

\rownumber & 
\href{http://skyserver.sdss.org/dr14/en/tools/explore/summary.aspx?sid=5386312396405972992&apid=}{SDSS J140237.96+034231.7}  \\

\rownumber & 
\href{http://skyserver.sdss.org/dr14/en/tools/explore/summary.aspx?sid=6534893589381488640&apid=}{SDSS J091337.33+295958.4}  \\

\rownumber & 
\href{http://skyserver.sdss.org/dr14/en/tools/explore/summary.aspx?sid=1780176998165932032&apid=}{SDSS J154024.75+325157.2}  \\

 \bottomrule
\end{tabular}
\caption{Galaxies showing type Ia supernova features in their spectra.}
\label{tab:supernova}

\large
\end{center}
\end{table}

%% file: isolated.txt
%\newcounter{magicrownumbers}
%\newcommand\rownumber{\stepcounter{magicrownumbers}\arabic{magicrownumbers}}

\setcounter{magicrownumbers}{0}

\begin{table}
\begin{center}

\tiny
\begin{tabular}{l|ccc}

\toprule
index & SDSS name  & Comments     \\
\midrule
\rownumber &  \href{http://skyserver.sdss.org/dr14/en/tools/explore/summary.aspx?sid=1953551013378025472&apid=}{SDSS J073714.26+431414.9} &  shifted set of absorption lines \\

\rownumber &  
\href{http://skyserver.sdss.org/dr14/en/tools/explore/summary.aspx?sid=7575260289722982400&apid=}{SDSS J152613.25+495322.5} &  shifted set of absorption lines \\

\rownumber & 
\href{http://skyserver.sdss.org/dr14/en/tools/explore/summary.aspx?sid=3636666126487887872&apid=}{SDSS J083316.31+152314.6} &  shifted set of absorption lines \\
 
 \rownumber & 
\href{http://skyserver.sdss.org/dr14/en/tools/explore/summary.aspx?sid=683536182147049472&apid=}{SDSS J142812.98+611115.6} &  extreme blueshift \nad{} \\

 \rownumber & 
\href{http://skyserver.sdss.org/dr14/en/tools/explore/summary.aspx?sid=889475809172023296&apid=}{SDSS J143815.47+570445.1} & chance alignment, brown dwarf \\

 \rownumber & 
\href{http://skyserver.sdss.org/dr14/en/tools/explore/summary.aspx?sid=8175252517975728128&apid=}{SDSS J052223.70+005916.4} & unknown \\

 \rownumber & 
\href{http://skyserver.sdss.org/dr14/en/tools/explore/summary.aspx?sid=2032305746833598464&apid=}{SDSS J135124.77+054903.1} & bad redshift \\

 \bottomrule
\end{tabular}
\caption{Galaxies that are isolated on the \nad{} \texttt{UMAP}, suggesting unique properties, not similar to any other galaxy in the sample.}
\label{tab:isolated}

\large
\end{center}
\end{table}